\documentclass[reqno]{amsart}
\usepackage[foot]{amsaddr}
\usepackage[UKenglish]{babel}
\usepackage[utf8]{inputenc}

\usepackage{microtype}
\usepackage{relsize}

\usepackage[textsize=footnotesize]{todonotes}

\usepackage{import}
\usepackage{pdfpages}
\usepackage{transparent}
\usepackage{xcolor}

\usepackage{amsthm}
\usepackage{amsmath}
\usepackage{amssymb}
\usepackage{booktabs}
\usepackage{stmaryrd}
\usepackage{faktor}

\usepackage{upgreek}

\usepackage{mathtools}
\usepackage{thmtools}
\usepackage{thm-restate}

\usepackage{tikz}
\usetikzlibrary{positioning}
\usepackage{tikz-cd}

\newcommand\defined{\triangleq}

\newcommand\psx{pre-spectral spaces}

\newcommand\Ps{Pre-spectral space}
\newcommand\ps{pre-spectral space}
\newcommand\lpps{lpps}

\newcommand\ltrsk{\L{}o\'s-Tarski Theorem}

\newcommand\EPFO{\mathsf{EPFO}}

\newcommand\PFO{\mathsf{EPFO}^{\neq}}
\newcommand\EFO{\mathsf{EFO}}
\newcommand\FO{\mathsf{FO}}

\newcommand\Mod{\operatorname{Struct}}
\newcommand\ModF{\operatorname{Fin}}
\newcommand\ModFMP{\operatorname{FMP}}

\newcommand\Diag{\operatorname{Diag}}

\newcommand\Td{\mathcal{T}}
\newcommand\td{\operatorname{td}}

\newcommand\Graphs{\mathcal{G}}%
\newcommand\Cycles{\mathcal{C}}%
\newcommand\sigmaG{\upsigma_{\mathcal{G}}}

\newcommand\homomorphism{\to}

\newcommand{\NCore}[2]{\operatorname{Core}^{#1}(#2)}
\newcommand\subfin{\subseteq_\mathrm{fin}}

\newcommand{\Alex}[1]{\uptau_{#1}}

\newcommand{\modset}[1]{\llbracket #1 \rrbracket}

\newcommand{\Sobrif}[1]{\mathcal{S}\left(#1\right)}

\newcommand{\LPS}[3]{\left\langle #1, #2, #3 \right\rangle}

\newcommand{\Rel}[2]{\mathbf{#1}^{\!#2}}

\newcommand{\CompSat}[1]{\mathcal{K}^{\circ}\!\left(#1\right)}

\newcommand{\CountableUnions}{\ModF^\uplus}
\newcommand{\Age}{\operatorname{Age}}

\newcommand\CatPS{\mathbf{PreSpec}}
\newcommand\CatSpec{\mathbf{Spec}}
\newcommand\CatTop{\mathbf{Top}}

\newcommand{\setof}[2]{\left\{ #1 \mid #2 \right\}}
\newcommand{\parts}[1]{\wp(#1)}

\usepackage{environ}
\NewEnviron{killcontents}{}

\newenvironment{sendappendix}{}{}
\newenvironment{ifappendix}{\killcontents}{\endkillcontents}
\newif\ifsubmission
\submissionfalse
\newcommand\appref\cref

\usepackage{url,hyperref}

\usepackage[numbers]{natbib}
\providecommand{\doi}[1]{\href{http://dx.doi.org/#1}{\nolinkurl{doi:#1}}}

\usepackage[capitalise,noabbrev,nameinlink]{cleveref}
\usepackage{thmtools,thm-restate}
\theoremstyle{plain}
\newtheorem{theorem}{Theorem}[section]
\newtheorem{lemma}[theorem]{Lemma}
\newtheorem{proposition}[theorem]{Proposition}

\newtheorem{fact}[theorem]{Fact}
\theoremstyle{definition}
\newtheorem{definition}[theorem]{Definition}
\newtheorem{example}[theorem]{Example}
\theoremstyle{remark}
\newtheorem{remark}[theorem]{Remark}
\newtheorem{claim}[theorem]{Claim}
\newenvironment{claimproof}[1][]{\begin{proof}[#1]}{\end{proof}}
\crefname{fact}{Fact}{facts}
\crefname{proposition}{Proposition}{propositions}
\Crefname{fact}{Fact}{Facts}
\crefname{enumi}{item}{items}
\Crefname{enumi}{Item}{Items}
\crefname{claim}{Claim}{claims}
\crefname{definition}{Definition}{definitions}
\crefname{lemma}{Lemma}{lemmas}
\renewcommand{\subparagraph}[1]{\subsubsection{#1\nopunct}}
\title[Preservation Theorems Through the Lens of
  Topology]{Preservation Theorems Through the Lens \mbox{of Topology}}
\author{Aliaume Lopez}
\thanks{%
    I would like to thank Denis Kuperberg
    and Quentin Moreau for spotting
    a mistake in 
    \cref{cor:prespectral:rhpt} regarding the fragment
    $\mathsf{NFO}$.
}
\address{Universit\'e Paris-Saclay, ENS Paris-Saclay, CNRS, LSV, France}
\email{aliaume.lopez@ens-paris-saclay.fr}
\begin{document}

    \begin{abstract}
      In this paper, we introduce a family of topological spaces that
captures the existence of preservation theorems.  The structure of
those spaces allows us to study the relativisation of preservation
theorems under suitable definitions of surjective morphisms,
subclasses, sums, products, topological closures, and projective
limits.  Throughout the paper, we also integrate already known results
into this new framework and show how it captures the essence of their
proofs.

    \end{abstract}
    \maketitle

    \section{Introduction}
    \label{sec:intro}

    In classical model theory, \emph{preservation theorems} characterise
first-order definable sets enjoying some semantic property as those
definable in a suitable syntactic fragment
\cite[e.g.,][Section~5.2]{chang1990model}.  A well-known instance is
the \L{}o\'s-Tarski Theorem~\cite{tarski54,los55}: a first-order
sentence~$\varphi$ is preserved under extensions on all structures---i.e.,
$A\models\varphi$ and $A$ is an induced substructure of~$B$ imply
$B\models\varphi$---if and only if it is equivalent to an existential
sentence.

A major roadblock for applying these results in computer science is
that preservation theorems generally do not relativise to classes of
structures, and in particular to the class of all finite
structures~[see the discussions in \citenum{rosen02}, Section~2
and \citenum{kolaitis07}, Section~3.4].  In fact, the only case where
a classical preservation theorem was shown to hold on all finite
structures is Rossman's Theorem~\cite{Rossman08}: a first-order
sentence is preserved under homomorphisms on all finite structures if
and only if it is equivalent to an existential positive sentence.
This long-sought result has applications in database theory, where
existential positive formul\ae\ correspond to unions of conjunctive
queries (also known as select-project-join-union queries and arguably
the most common database queries in
practice~\cite{abiteboul1995foundations}).  For instance, it is
related in~\cite[Theorem~17]{DeutschNR08} to the existence of
homomorphism-universal models (as constructed by chase algorithms) for
databases with integrity constraints, in~\cite[Theorem~3.4]{tencate09}
to a characterisation of schema mappings definable via
source-to-target tuple-generating dependencies, and
in~\cite[Corollary~4.14]{gheerbrant2014naive} to the na\"ive
evaluation of queries over incomplete databases under open-world
semantics.  These applications would benefit directly from
preservation theorems for more restricted classes of finite structures
or for other semantic properties---corresponding to other classes of
queries and other semantics of incompleteness---, and this has been an
active area of
research~\cite{atserias2006preservation,atserias2008preservation,dawar2010homomorphism,harwath2014preservation,figueira14}.
Like \citeauthor{Rossman08}'s result, these proofs typically rely on
careful model-theoretic arguments---typically using
Ehrenfeucht-Fra\"isse games and locality---and each new attempt at
proving a preservation theorem seemingly needs to restart from
scratch.

\medskip
In this paper, we develop a general topological framework for
investigating preservation theorems, where preservation theorems, both
old and new, can be obtained as byproducts of topological
constructions.

As pointed out in the literature, the classical proofs of preservation
theorems fail in the finite because the Compactness Theorem does not
apply.  As we will see in \cref{sec:preservation}, one can reinterpret
in topological terms the two applications of the Compactness Theorem
in the classical proofs of preservation theorems like the
\L{}o\'s-Tarski Theorem.  Here, the topology of interest has the sets
of structures closed under extension as its open sets, and one
application of the Compactness Theorem shows that the definable open
sets are compact~(\cref{lem:classical:compactness}) while the other
shows that the sets definable by existential sentences form a base for
the definable open sets~(\cref{lem:classical:generating}).  In
\cref{sec:pspec}, we capture these two ingredients in general through
the definitions of \emph{logically presented pre-spectral spaces}
and \emph{diagram bases} in \cref{sec:pspec}, which lead to a generic
preservation theorem (\cref{thm:prespectral:rhpt}): under mild
hypotheses---which are met in all the preservation results over
classes of finite structures in the literature---, preservation holds
if and only if the space under consideration is logically presented
pre-spectral.
\medskip

The benefit of this abstract, topological viewpoint, is that
preservation results can now be proven by constructing new logically
presented pre-spectral spaces from known ones.

Here, the topological core of our definition is the one
of \emph{pre-spectral} spaces, which generalise both Noetherian spaces
and spectral spaces~\cite{goubault2013non,spectral2019};
see \cref{sec:related}.  To some extent, we can rely on the stability
of spectral spaces under various topological constructions to
investigate the same constructions for pre-spectral spaces.  We focus
however in the paper on the \emph{logically presented} pre-spectral
spaces, which is where the main difficulty lies when attempting to
prove preservation over classes of finite structures, and for which
stability must take the logical aspect into account.
Accordingly, \cref{sec:closure} shows the stability of logically
presented pre-spectral spaces under typical constructions: under a
carefully chosen notion of morphisms, under subclasses provided a
sufficient condition is met, and under finite sums and finite
products.

Where the topological viewpoint really shines is when it comes to
stability for various kinds of `limits' of classes of structures
enjoying a preservation property.  We show in \cref{sec:logic} that
the limit of a \emph{single} class of structures, when it can be
construed as the \emph{closure} in a suitable topology of a logically
presented pre-spectral space, is also logically presented
pre-spectral.  This allows us to show that \citeauthor{Rossman08}'s
Theorem---i.e., homomorphism preservation in the finite---extends to
the class of structures with the finite model property, and also
extends to countable unions of finite structures (the latter was also shown
in~\cite[Chapter~10]{Neetil12}).  In \cref{sec:proj}, we show that the
limit of a \emph{family} of pre-spectral spaces, when built as a
\emph{projective limit}, is also pre-spectral.  We use this to show
that \citeauthor{Rossman08}'s proof of homomorphism preservation in
the finite can be re-cast in our framework as building exactly such a
projective limit.

\ifsubmission Due to space constraints, detailed proofs and additional
examples will be found in the appendices.\fi

    \section{Preservation Theorems}
    \label{sec:preservation}
    In this section, we revisit classical preservation theorems,
whose proofs can be found in many books such
as~\cite[Section~5.2]{chang1990model}.  We will recall the needed
definitions, and illustrate the proof techniques in order
to highlight the two ingredients that motivate our
definitions of {\psx} and diagram bases later in \cref{sec:pspec}.

\subsection{Classical Preservation Theorems}

\subparagraph{Notations.}
A $\upsigma$-structure~$A$ over a finite relational
signature~$\upsigma$ (without constants) is given by a domain~$|A|$
and, for each symbol $R \in \upsigma$ of arity~$n$, a
relation~$\mathbf{R}^A\subseteq|A|^n$; $A$ is finite if $|A|$ is
finite.  The binary symbol~`$=$' will always be interpreted as
equality, and will not be explicitly listed in our signatures. We
write $\Mod(\upsigma)$ for the set\footnote{In order to work over sets
instead of proper classes and thereby avoid delicate but out-of-topic
foundational issues, every %
$\upsigma$-structure in this paper will be assumed to be of
cardinality bounded by some suitable infinite cardinal%
.  In
particular, the L\"owenheim-Skolem Theorem justifies that this is at
no loss of generality when working with first-order logic.} of all the
$\upsigma$-structures and $\ModF(\upsigma)$ for the finite ones.
We assume the reader is familiar with the syntax and semantics of
first-order logic over~$\upsigma$.  We write $\FO[\sigma]$ for the set
of first-order sentences over~$\upsigma$. For such a
sentence~$\varphi$, we write $\modset{\varphi}_X \defined \left\{
A \in X \mid A \models \varphi \right\}$ for its set of models over a
class of structures~$X\subseteq\Mod(\upsigma)$; by extension, we let
$\modset{\mathsf{F}}_X\defined\{\modset{\varphi}_X\mid\varphi\in\mathsf{F}\}$
denote the collection of $\mathsf{F}$-definable subsets of~$X$ for a
fragment $\mathsf{F}$ of $\FO[\sigma]$.

\subparagraph{Abstract Preservation.}
A preservation theorem over a class of
structures~$X\subseteq\Mod(\upsigma)$ shows that first-order sentences
enjoying some semantic property are equivalent to sentences from a
suitable a syntactic fragment.  More precisely, one can model a
semantic property as a collection~$\mathcal O\subseteq\parts X$ of
`semantic observations' and consider a fragment~$\mathsf
F\subseteq\FO[\sigma]$: we will say that $X$ has the
\emph{$(\mathcal{O},\mathsf{F})$ preservation property} if
\begin{enumerate}
\item\label{pres1} for all $\psi\in\mathsf F$, $\modset{\psi}_X\in\mathcal
  O$%
  , and,
\item\label{pres2} for all $\varphi\in\FO[\sigma]$ such that
  $\modset{\varphi}_X\in\mathcal{O}$, there exists~$\psi\in\mathsf{F}$
  such that $\modset{\varphi}_X=\modset{\psi}_X$%
  .
\end{enumerate} In this definition, \cref{pres1} is
usually proven by a straightforward induction on the formul\ae\
in~$\mathsf F$, and the challenge is to
establish \cref{pres2}.  \Cref{pres2} is also
where \emph{relativisation} to a subset $Y\subseteq X$ might fail,
because a set $U\not\in\mathcal O$ might still be such that $U\cap
Y\in\{V\cap Y\mid V\in\mathcal O\}$, and thus there might be new
first-order sentences enjoying the semantic property and requiring an
equivalent sentence in~$\mathsf F$.

Put more succinctly, $X$ has the $(\mathcal{O},\mathsf{F})$
preservation property if
\begin{equation}
\label{abs-pres}
  \mathcal{O}\cap \modset{\FO[\upsigma]}_X = \modset{\mathsf{F}}_X\;.
\end{equation}
This formulation explicitly shows how a semantic condition (the
left-hand side in~\eqref{abs-pres}) is matched with a syntactic one
(the right-hand side).  As preservation is of interest beyond
first-order logic~\cite[e.g.,][]{grohe97,feder03,dawar08,figueira14},
we will say in full generality that a set~$X$ equipped with a lattice
$\mathcal{L}$ of sets definable in the logic of interest has
the \emph{$(\mathcal{O},\mathcal{L}')$ preservation property} if
\begin{equation}
    \mathcal{O}\cap \mathcal{L} = \mathcal{L}'
\end{equation}

In the rest of this paper
we will assume
that $\mathcal{O}$ contains~$\emptyset$, contains~$X$, is closed
under finite intersections and arbitrary unions.
This is equivalent to
$\mathcal{O}$ being a collection of open sets and defining
a \emph{topology} on~$X$.

\subparagraph{Monotone Preservation.}
In a number of cases, which are especially relevant in the
applications to database theory mentioned in the
introduction~\cite{DeutschNR08,gheerbrant2014naive}, the semantic
property of interest is a form of monotonicity for some quasi-ordering
$\leq$ of $\Mod(\upsigma)$.  We say that a sentence $\varphi$
is \emph{monotone} in~$X\subseteq\Mod(\upsigma)$ if
$\modset{\varphi}_X$ is \emph{upwards-closed}, meaning that if
$A\in\modset{\varphi}_X$ and $B$ is a $\upsigma$-structure in~$X$ such
that $A\leq B$, then $B\in \modset{\varphi}_X$.  In terms of abstract
preservation, this corresponds to choosing $\mathcal O$ as the
collection of upwards-closed subsets of~$X$, which is also known as
the \emph{Alexandroff topology} and is denoted by~$\uptau_\leq$.

\begin{table}
    \caption{Classical preservation theorems and their relativisations to the
        finite case.}
    \label{table:classical:cpres}
    \centering
    \begin{tabular}{lccl}
        \toprule
        \textbf{preservation theorem} & \textbf{\boldmath quasi-ordering~$\leq$}
        & \textbf{\boldmath fragment~$\mathsf{F}$} & \textbf{\boldmath holds in~$\ModF(\upsigma)$} \\
        \midrule
        homomorphism & $\homomorphism$ & $\EPFO$
        & yes~\cite{Rossman08} \\
        Tarski-Lyndon & $\subseteq$ & $\PFO$  &  no~\cite{ajtai1994datalog} \\
        \L{}o\'s-Tarski & $\subseteq_i$ & $\EFO$ &
        no~\cite{tait1959counterexample,gurevitch84,dawar20} \\
        dual Lyndon & $\twoheadleftarrow$ & $\mathsf{NFO}$ &
        no~\cite{ajtai87,stolboushkin95}\\
        \bottomrule
    \end{tabular}
\end{table}

The quasi-ordering~$\leq$ in question is typically defined through
some class of homomorphisms.  Recall that there is
a \emph{homomorphism} between two $\upsigma$-structures~$A$ and~$B$,
noted $A\to B$, if there exists $f\colon |A| \to |B|$ such that, for
all relation symbols $R$ of~$\upsigma$ and all tuples
$(a_1,\dots,a_n)\in\Rel{R}{A}$, $(f(a_1), \dots,
f(a_n))\in\Rel{R}{B}$.  When $f$ is injective, this entails that $A$
is (isomorphic to) a (not necessarily induced)
\emph{substructure} of~$B$ and we write $A \subseteq B$; when $f$ is
furthermore \emph{strong}---meaning that for all~$R$ and
$(a_1,\dots,a_n)\in|A|^n$, $(f(a_1), \dots, f(a_n))\in\Rel{R}{B}$
implies $(a_1,\dots,a_n)\in\Rel{R}{A}$---, this entails that $A$ is
(isomorphic to) an \emph{induced substructure} of~$B$ and we
write~$A\subseteq_i B$; finally, we write $A\twoheadrightarrow B$ when
$f$ is surjective%
.

\Cref{table:classical:cpres} summarises what is known about monotone
preservation theorems.  In this table, $\EFO$ denotes the set of
existential first-order sentences, $\mathsf{NFO}$ the set of negative
ones (namely negative atoms closed under~$\vee$, $\wedge$, $\exists$,
and~$\forall$), $\EPFO$ the set of existential positive ones, and
$\PFO$ the set of existential positive ones extended with atoms of the
form $x \neq y$ (interpreted as inequality)%
.  Note that Lydon's Theorem, which states that a first-order sentence
closed under surjective homomorphisms on all structures is equivalent
to a positive one, is presented in \cref{table:classical:cpres} in
its \emph{dual} form with inverse surjective homomorphisms and negative
sentences.  For all these fragments~$\mathsf F$ and associated
quasi-orderings~$\leq$, the fact that $\modset{\mathsf
F}_X\subseteq\uptau_\leq$ is mostly straightforward.

\subsection{The \ltrsk\ in Topological Terms}
\label{sub:ltrsk}
We propose now to inspect the proof of the {\ltrsk} on a finite
relational signature~$\upsigma$, as found for instance
in~\cite[Theorem~3.2.2]{chang1990model}
or \cite[Section~5.4]{hodges1997shorter}.  We work here with the
collection $\mathcal O\defined\uptau_{\subseteq_i}$ of upwards-closed
subsets of $X\defined\Mod(\upsigma)$ for~$\subseteq_i$ (this is the
Alexandroff topology of the quasi-order~$\subseteq_i$) and the
fragment $\mathsf F\defined\EFO[\upsigma]$.  The {\ltrsk} corresponds
to the following instantiation of \eqref{abs-pres}:
\begin{equation}
    \label{eqn:classical:ltrsk} \uptau_{\subseteq_i} \cap \modset{\FO[\upsigma]}_{\Mod(\upsigma)}
    = \modset{\EFO[\upsigma]}_{\Mod(\upsigma)}\;.
\end{equation}
The proof of the {\ltrsk} can be decomposed into two steps, here
corresponding to the
upcoming \cref{lem:classical:generating,lem:classical:compactness},
and each invoking the Compactness Theorem.  When translated in
topological terms, the first shows that $\EFO$ defines a \emph{base}
for the definable open sets, while the second shows that definable
open sets are \emph{compact}.

\subparagraph{`Syntactic' Base.}
Recall that a \emph{base} $\mathcal{B}$ of a topology~$\uptau$ is a
collection of open sets such that every open set of~$\uptau$ is a
(possibly infinite) union of elements
from~$\mathcal{B}$. Equivalently, $\mathcal{B}$ is a base of a
topology~$\uptau$ whenever $\forall U \in \uptau,\forall A \in U,
\exists V \in \mathcal{B}, A \in V \subseteq U$.
A \emph{subbase} is a collection of open sets such that every open set
of~$\uptau$ is a (possibly infinite) union of finite intersections of
elements of the \emph{subbase}. The topology
$\langle \mathcal{O} \rangle$
\emph{generated} by a collection $\mathcal{O}$ of sets is the smallest
topology containing those sets; $\mathcal{O}$ is then a subbase of
$\langle \mathcal{O} \rangle$.

We first prove a weaker version of~\cref{eqn:classical:ltrsk} by
proving the equality on the generated topologies.  Because
$\modset{\FO[\upsigma]}_{\Mod(\upsigma)}$
and $\modset{\EFO[\upsigma]}_{\Mod(\upsigma)}$
are lattices, those generated
topologies can be seen as generated by infinite disjunctions of
sentences in $\FO[\upsigma]$ (resp.\ $\EFO[\upsigma]$).

\begin{claim}%
    \label{lem:classical:generating}
  The topologies generated by
  $\uptau_{\subseteq_i} \cap \modset{\FO[\upsigma]}_{\Mod(\upsigma)}$
  and $\modset{\EFO[\upsigma]}_{\Mod(\upsigma)}$ are the same, i.e.,
            $\left\langle \uptau_{\subseteq_i} \cap \modset{\FO[\upsigma]}_{\Mod(\upsigma)} \right\rangle
            =
        \left\langle \modset{\EFO[\upsigma]}_{\Mod(\upsigma)}\right\rangle$.
\end{claim}
\begin{claimproof}
    First of all, any sentence in $\EFO[\upsigma]$ defines an
    upwards-closed set for~$\subseteq_i$, and moreover
    $\EFO[\upsigma] \subseteq \FO[\upsigma]$, hence
    $\left\langle \modset{\EFO[\upsigma]}_{\Mod(\upsigma)} \right\rangle\subseteq\left\langle \uptau_{\subseteq_i} \cap \modset{\FO[\upsigma]}_{\Mod(\upsigma)} \right\rangle$.

    For the converse inclusion, it suffices to show that
    $\EFO[\upsigma]$ defines a base of the topology
    $\langle \tau_{\subseteq_i} \cap \modset{\FO[\upsigma]}_{\Mod(\upsigma)} \rangle$.
    Consider for this a monotone sentence $\varphi \in \FO[\upsigma]$
    and a structure~$A$ such that $A \models \varphi$.  Following the
    classical proofs~[e.g., \citenum{chang1990model}, Theorem 3.2.2
    or \citenum{hodges1997shorter}, Corollary~5.4.3], define $\hat{A}$
    as the expansion of~$A$ with one additional constant~$c_a$ for
    each~$a\in|A|$, interpreted by $c^{\hat A}_a \defined
    a$. The \emph{diagram} $\Diag(A)$ of~$A$ is the set of all
    quantifier-free sentences over this extended signature that hold
    in~$\hat A$.  For a structure
    $\hat{B} \in \Mod(\upsigma\cup\{c_a\}_{a \in A})$, we write~$B$
    for its reduct in $\Mod(\upsigma)$ obtained by removing the
    constants~$\{c_a\}_{a \in A}$.

    Let $T \defined \Diag(A) \cup \{ \neg \varphi \}$, and consider
    $\hat{B} \in \Mod(\upsigma\cup\{c_a\}_{a \in A})$ such that
    $\hat{B} \models T$.  Because $\hat{B} \models \Diag(A)$, by
    construction $A \subseteq_i B$ (in particular, the sentence
    $\neg(c_a=c_b)$ belongs to $\Diag(A)$ for all $a\neq b$ in~$|A|$),
    and thus $B \models \varphi$ because $\varphi$ is monotone, and
    finally $\hat{B} \models\varphi$ because the constants~$c_a$ do
    not occur in~$\varphi$.  Therefore, $\hat
    B \models \varphi \wedge \neg \varphi$, which is absurd: the
    theory~$T$ is inconsistent, and by the Compactness Theorem for
    first-order logic, there exists a finite conjunction~$\psi_0$
    of sentences in $\Diag(A)$, which is already inconsistent with
    $\neg \varphi$.

    Let~$\psi_A$ be the existential closure of the formula obtained by
    replacing each symbol~$c_a$ with a variable~$x_a$ in~$\psi_0$;
    note that~$\psi_A$ is an existential sentence.
    By construction, $A \models \psi_A$, and if
    $B \models \psi_A$, there exists an interpretation of the
    constants $\{c_a\}_{a \in A}$ allowing to build an
    expansion~$\hat{B}$ such that $\hat{B} \models \psi_0$.  As we
    just saw that $\models \psi_0 \implies \varphi$,
    $\hat{B} \models \varphi$, and since no constant symbol occurs in
    $\varphi$, $B \models \varphi$.

    To conclude, for any open set
    $U\in\langle\tau_{\subseteq_i} \cap \modset{\FO[\upsigma]}_{\Mod(\upsigma)}\rangle$
    and for any $A \in U$, there exists a monotone sentence $\varphi$ such
    that $A \in\modset{\varphi}_{\Mod(\upsigma)}$, and we have proven
    that there exists
    $\modset{\psi_A}_{\Mod(\upsigma)} \in \modset{\EFO[\upsigma]}_{\Mod(\upsigma)}$
    such that
    $A \in \modset{\psi_A}_{\Mod(\upsigma)} \subseteq \modset{\varphi}_{\Mod(\upsigma)}\subseteq
    U$.  Therefore, $\modset{\EFO[\upsigma]}_{\Mod(\upsigma)}$ is a
    base of
    $\langle\tau_{\subseteq_i} \cap \modset{\FO[\upsigma]}_{\Mod(\upsigma)}\rangle$.
\end{claimproof}

\subparagraph{Compactness.}
The second step relies on the compactness of the sets
$\modset{\varphi}_{\Mod(\upsigma)}$ for monotone sentences~$\varphi$.
Recall that a subset $K$ is \emph{compact} in a topological
space~$\uptau$ if, for any
\emph{open cover} $(U_i)_{i \in I}$ of~$K$---i.e., a collection of open sets such that $K
\subseteq \bigcup_{i \in I} U_i$---, there exists a finite subset $I_0 \subseteq
I$, such that $K \subseteq \bigcup_{i \in I_0} U_i$ (beware that this
definition is also called \emph{quasi-compact} in the literature,
because we do not require any separation property here).
If $\uptau=\langle\mathcal O\rangle$, by \emph{Alexander's Subbase
Lemma}, $K$ is compact if and only if, from every open cover of $K$
using only sets from $\mathcal{O}$, we can extract a finite open cover
of~$K$.  As open compact sets play a key role in this paper, we
introduce here the notation $\CompSat{X} \defined \{ U \in \uptau\mid
U\text{ is compact}\}$.  When the topology~$\uptau$ is not clear from
the context, we shall write $\CompSat{X, \uptau}$.%

\begin{claim}%
    \label{lem:classical:compactness}
    Every monotone sentence defines a compact open subset in the
    topology $\langle\uptau_{\subseteq_i}\cap \modset{\FO[\upsigma]}_{\Mod(\upsigma)}\rangle$, i.e.,
        $\uptau_{\subseteq_i} \cap \modset{\FO[\upsigma]}_{\Mod(\upsigma)} \subseteq \CompSat{\Mod(\upsigma)\ifsubmission,\langle\uptau_{\subseteq_i}\cap \modset{\FO[\upsigma]}_{\Mod(\upsigma)}\rangle\fi}$.
\end{claim}

\begin{claimproof}
    Consider a monotone sentence
    $\varphi \in \FO[\upsigma]_{\Mod(\upsigma)}$.  Let $(U_i)_{i \in
    I}$ be an open cover of $\modset{\varphi}_{\Mod(\upsigma)}$.  By
    Alexander's Subbase Lemma, we can assume that for each $i\in I$,
    $U_i = \modset{\varphi_i}_{\Mod(\upsigma)}$ for some monotone
    sentence $\varphi_i$.  Consider the theory
    $T \defined \{ \neg \varphi_i \mid i \in
    I \} \cup \{ \varphi \}$. Because $(U_i)_{i \in I}$ is an open
    cover, this theory has no models.  By the Compactness Theorem for
    first order logic, there exists a finite set $I_0$ such that
    $T_0 \defined \{ \neg \varphi_i \mid i \in
    I_0 \} \cup \{ \varphi \}$ is not satisfiable, proving that
    $(U_i)_{i \in I_0}$ is an open cover of
    $\modset{\varphi}_{\Mod(\upsigma)}$.
\end{claimproof}
\begin{remark}[Compact sets in $\Alex{\leq}$]
  \label{rk:alex:compact} As we will often deal %
  with the Alexandroff topology~$\Alex{\leq}$ of a
  quasi-order~$(X,{\leq})$, it is worth noting that
  $U\in\Alex{\leq}$ is compact if and only if it is the upward closure
  $U={\uparrow}F$ of some finite subset~$F\subfin X$; this is
  equivalent to saying that~$U$ has finitely many minimal elements up
  to
  $\leq$-equivalence~\citep[see e.g.,][Exercise~4.4.22]{goubault2013non}.
  Thus \cref{lem:classical:compactness} states that any monotone
  sentence has finitely many $\subseteq_i$-minimal models
  in~$\Mod(\upsigma)$.
\end{remark}

\begin{proof}[\ifsubmission\relax\else \textnormal{2.2.3.} \fi Proof of the {\ltrsk}]
    A simple structural induction on the formul\ae\ shows that
    $\modset{\EFO[\upsigma]}_{\Mod(\upsigma)} \subseteq \Alex{\subseteq_i} \cap \modset{\FO[\upsigma]}_{\Mod(\upsigma)}$.
    Regarding the converse inclusion in~\cref{eqn:classical:ltrsk},
    consider a sentence $\varphi \in \FO[\upsigma]$ defining an open
    set in $\Alex{\subseteq_i}$.  By~\cref{lem:classical:generating},
    there exists a family $(\psi_i)_{i \in I}$ of existential
    sentences such that $\modset{\varphi}_{\Mod(\upsigma)}
    = \bigcup_{i \in I} \modset{\psi_i}_{\Mod(\upsigma)}$.
    By~\cref{lem:classical:compactness}, there is a finite set
    $I_0 \subfin I$ for which the equality still holds.  Because
    $\EFO[\upsigma]$ is a lattice, this proves the existence of an
    existential sentence $\psi\defined\bigvee_{i\in I_0}\psi_i$ such
    that $\modset{\varphi}_{\Mod(\upsigma)}
    = \modset{\psi}_{\Mod(\upsigma)}$.
\end{proof}

The two properties singled out
in \cref{lem:classical:generating,lem:classical:compactness} are of
different nature.  \Cref{lem:classical:compactness} really holds for
any topology~$\uptau$ and not only for the Alexandroff topology
$\Alex{\subseteq_i}$, as opposed to~\cref{lem:classical:generating}.
Moreover, \cref{lem:classical:generating} appears to be the most
involved one here, but is often easily proven on classes of finite
structures.  %

    \section{Pre-spectral Spaces and Diagram Bases}
    \label{sec:pspec}
    Following the two-step decomposition of the proof of the \ltrsk\ given
in \cref{sub:ltrsk}, we define in this section \emph{logically
presented pre-spectral spaces} and \emph{diagram bases}, before
showing in \cref{thm:prespectral:rhpt} how they characterise when a
preservation theorem holds.

\subsection{Pre-spectral Spaces}
\label{sec:diagrambase}
As a preliminary step toward our main definition, let us first propose
a definition of topological spaces~$(X,\uptau)$ where the compact open
sets form a \emph{bounded sublattice} of $\parts X$ (by which we mean
that~$\emptyset$ and~$X$ belong to the lattice) that generates the
topology.
\begin{definition}[\protect\Ps]
    A topological space $(X,\uptau)$ is a \emph{\ps} whenever
    $\CompSat{X}$ is a bounded sublattice of $\parts X$ that
    generates~$\uptau$, i.e., $\langle \CompSat{X} \rangle = \uptau$.
\end{definition}
The name `pre-spectral' comes from the theory of
\emph{spectral} spaces~\cite{spectral2019}, for which the definition
is almost identical (see \cref{sec:spectral}).  Pre-spectral spaces
will allow us to tap into the rich topological toolset that has been
developed for spectral spaces.

\subparagraph{Logical presentations.} As seen
in \cref{lem:classical:compactness}, the topology of interest in a
preservation theorem is generated by combining a topological
space~$(X,\uptau)$ with a bounded sublattice~$\mathcal{L}$ of subsets
of~$X$, which will be called the \emph{definable} subsets of~$X$.  Let
us write $\LPS{X}{\uptau}{\mathcal{L}}$ for the topological space
$(X, \langle \uptau \cap \mathcal{L} \rangle)$.  The following
definition is then a direct generalisation of the statement
of \cref{lem:classical:compactness}.

\begin{definition}[Logically presented \ps]
  \label{def:prespectral:lps} Let $(X,\uptau)$ be a topological space
  and $\mathcal{L}$ be a bounded sublattice of $\parts X$.  Then
  $\LPS{X}{\uptau}{\mathcal{L}}$ is a \emph{logically presented \ps}
  (a \emph{\lpps}) if %
  its definable open subsets are compact, i.e., if
  $\uptau \cap \mathcal{L} \subseteq \CompSat{X}$.
\end{definition}
Whenever $\upsigma$ is a finite relational signature,
$X \subseteq \Mod(\upsigma)$ for a topological space~$(X,\uptau)$ and
$\mathcal L=\modset{\FO[\upsigma]}_X$, we %
denote it by $\LPS{X}{\uptau}{\FO[\upsigma]}$ for simplicity; e.g.,
$\LPS{\Mod(\upsigma)}{\Alex{\subseteq_i}}{\FO[\upsigma]}$ is a \lpps\ 
by~\cref{lem:classical:compactness}.

As $\uptau \cap \mathcal{L}$ is closed under finite intersection, any
open set in $\langle\uptau \cap \mathcal{L}\rangle$ is a union of sets
from $\uptau \cap \mathcal{L}$, thus any compact open set in
$\CompSat{X}$ is a finite union of sets from
$\uptau \cap \mathcal{L}$.  As $\uptau \cap \mathcal{L}$ is also
closed under finite unions, this shows the inclusion
$\CompSat{X}\subseteq \uptau \cap \mathcal{L}$.  Thus, in a {\lpps},
$\CompSat{X}=\uptau \cap \mathcal{L}$ is a bounded lattice and any
{\lpps} is indeed a \ps.  Conversely, $\LPS{X}{\uptau}{\CompSat{X}}$
is well-defined whenever $(X, \uptau)$ is a pre-spectral space; in
this case $\LPS{X}{\uptau}{\CompSat{X}}$ is a {\lpps}
and it equals
$(X,\uptau)$ (they have the same points and opens).

Beware however that
$(X, \langle \uptau \cap \mathcal{L} \rangle)=\LPS{X}{\uptau}{\mathcal{L}}$
being pre-spectral does not entail that it is a \lpps;
see \cref{rk:prespectral:lpps} at the end of the section.  While
pre-spectral spaces capture the topological core
behind \cref{lem:classical:compactness} with a simple definition, the
logically presented ones are the real objects of interest as far as
preservation theorems are concerned, and most of the technical
difficulties arising in the remainder of the paper will be concerned
with those.

\subsection{Diagram Bases}
Regarding~\cref{lem:classical:generating}, we simply
turn the statement of the claim into a definition, which is typically
instantiated with $\mathcal L=\modset{\FO[\upsigma]}_X$ and $\mathcal
L'=\modset{\mathsf F}_X$ for a fragment $\mathsf F$ of $\FO[\upsigma]$.

\begin{definition}[Diagram base]
    \label{def:prespectral:diagbas}
    Let $(X,\uptau)$ be a topological space,
    and $\mathcal{L}$ be a bounded sublattice of $\wp(X)$.
    Then $\mathcal{L}'\subseteq\mathcal{L}$
    is a \emph{diagram
    base} of $\LPS{X}{\tau}{\mathcal{L}}$
    if
    $ \left\langle \uptau \cap \mathcal{L} \right\rangle
    = \left\langle \mathcal{L'} \right\rangle$.
\end{definition}
In particular, if $\mathsf F \subseteq \FO[\upsigma]$
is stable under finite conjunction,
this means that any definable open set in $X$
can be written as an infinite
disjunction of $\mathsf{F}$-definable sets.  Over $\Mod(\upsigma)$,
this was the `difficult' step in the classical proof of the {\ltrsk}.
When $X\subseteq \ModF(\upsigma)$, this becomes considerably simpler:
for every fragment~$\mathsf F$ in~\cref{table:classical:cpres} and any
finite structure~$A$, there exists a \emph{diagram} sentence~$\psi_A^\mathsf{F}$
in~$\mathsf F$ such that $A\leq B$ if and only if $B\models\psi_A^\mathsf{F}$ for
the corresponding quasi-ordering.  Therefore, if $\varphi$ is monotone
and $A\in \modset{\varphi}_X$, then
$A\in \modset{\psi_A^\mathsf{F}}_X\subseteq \modset{\varphi}_X$, showing
that $\modset{\mathsf{F}}_{X}$ is a base of
$\left\langle \Alex{\leq} \cap \modset{\FO[\upsigma]}_{X} \right\rangle$.

\subsection{A Generic Preservation Theorem}

We have already seen in the proof of the {\ltrsk} why logically
presented pre-spectral spaces with a diagram base yield preservation.
The following theorem also proves the converse direction, under mild
hypotheses on~$\mathcal L'$: $\mathcal L'$ must be a lattice and must
define compact sets in~$X$ for the topology generated by~$\mathcal
L'$.  We usually instantiate the theorem with
$X \subseteq \Mod(\upsigma)$ $\mathcal{L} = \modset{\FO[\upsigma]}_X$
and $\mathcal{L}' = \modset{\mathsf{F}}_X$ where $\mathsf{F}$ is a
fragment of~$\FO[\upsigma]$.

\begin{theorem}[Generic preservation]
    \label{thm:prespectral:rhpt} 
    Let $\uptau$ be a topology on $X$,
    $\mathcal{L}$ a bounded sublattice of $\wp(X)$,
    and $\mathcal{L}'$ a sublattice of $\mathcal{L}$.
    The following are equivalent:
    \begin{enumerate}
        \item\label{gen-pres1} $X$ has the $(\uptau,\mathcal L')$
    preservation property and $\mathcal{L}'$ defines only compact sets
    for the topology~$\langle\mathcal{L}'\rangle$.
    \item\label{gen-pres2} $\LPS{X}{\uptau}{\mathcal{L}}$ is a \lpps\ and
        $\mathcal{L}'$ is a diagram base of it.
    \end{enumerate}
\end{theorem}
\begin{proof}
    We prove the two implications separately.
    \begin{enumerate}
    \item Assume that $X$ has the $(\uptau,\mathcal L')$ preservation
        property. Consider a set $U \in \mathcal{L} \cap \uptau$:
        by the preservation property, $U \in \mathcal{L}'$.
        This already shows
        that $\mathcal{L}'$ defines a diagram base of
        $\LPS{X}{\uptau}{\mathcal{L}}$.
        Hence $\langle \mathcal{L}' \rangle =
        \langle \uptau \cap \mathcal{L} \rangle$.
        Since $U \in \mathcal{L}'$, $U$ is compact in
        $\langle \mathcal{L}' \rangle$, which means
        that $U$ is compact in $X$.
        Therefore $X$ is a \lpps.

    \item Assume that $\mathcal{L}'$ defines a diagram base of
        $\LPS{X}{\uptau}{\mathcal{L}}$.
        If $U \in \uptau \cap \mathcal{L}$,
        then it is equivalent to a possibly infinite
        union of elements in $\mathcal{L}'$.
       Also assume that
       $\LPS{X}{\uptau}{\mathcal{L}}$ is a \lpps: then by compactness,
    $U$ is equivalent to a finite union of elements in
    $\mathcal{L}'$, hence equivalent to a single element in $\mathcal{L}'$
    since $\mathcal{L}'$ is a lattice.
    This proves that $X$ has the $(\uptau,\mathcal L')$ preservation
    property. Finally, sets in $\mathcal{L}'$ define compact sets
    in $\langle\mathcal{L}'\rangle$ because it is
    precisely the topology of
    $\LPS{X}{\uptau}{\mathcal{L}}$.\qedhere \end{enumerate}
\end{proof}

The additional hypotheses on~$\mathcal L'$
in \cref{gen-pres1,gen-pres2} above are somewhat at odds.  Asking for
$\mathcal{L}'$ to define a diagram base is asking for
$\langle\mathcal{L}'\rangle$ to have enough sets, but asking for
$\mathcal{L}'$ to only define compact sets is asking for
$\langle\mathcal{L}'\rangle$ not to contain too many sets.

\begin{remark}[Generic monotone preservation]
    \label{cor:prespectral:rhpt} The condition that~$\mathsf F$ must
    define compact sets in~$X$
    in \cref{thm:prespectral:rhpt}.\ref{gen-pres1} is actually mild.
    Consider the preservation results
    from~\cref{table:classical:cpres} for a fragment $\mathsf F$ and
    $\uptau=\Alex{\leq}$ the Alexandroff topology of the associated
    quasi-ordering~$\leq$.  Assume that~$X$ is a
    $\leq$-downwards-closed subset of~$\Mod(\upsigma)$---this is the
    setting of the known preservation results for classes of finite
    structures~\cite{atserias2006preservation,atserias2008preservation,Rossman08,dawar2010homomorphism,harwath2014preservation}.

    Observe that, in each case \textbf{except for $\mathsf{NFO}$},
    $\modset{\psi}_{\Mod(\upsigma)}$ for a sentence~$\psi\in\mathsf F$ has
    finitely many $\leq$-minimal models up to $\leq$-equivalence.  Because $X$
    is downwards-closed, $\modset{\psi}_{X}$ has the same finitely many
    $\leq$-minimal models. Thus, by \cref{rk:alex:compact}, $\modset{\psi}_{X}$
    is compact in $\Alex{\leq}$, and since~$\modset{\mathsf
    F}_{X}\subseteq\Alex{\leq}$, it is also compact in the topology generated
    by~$\modset{\mathsf{F}}_X$.

    In the case of $X=\ModF(\upsigma)$, this downward
    closure condition is fulfilled and $\mathsf F$ defines a base,
    thus $(\Alex{\leq},\mathsf F)$ preservation holds if and only if
    $\LPS{\ModF(\upsigma)}{\Alex{\leq}}{\FO[\upsigma]}$ is a \lpps.
\end{remark}

\Cref{thm:prespectral:rhpt}
is a generic relationship between pre-spectral spaces and preservation
theorems. The downward closure hypothesis in
\cref{cor:prespectral:rhpt} is necessary for the equivalence between
the preservation property and pre-spectral spaces to hold, as will be
shown later in~\cref{ex:related:cycles}.
\begin{sendappendix}
\begin{ifappendix}\subparagraph{Downward Closures.}~\unskip\end{ifappendix}%
In order to apply~\cref{cor:prespectral:rhpt} to a set that
is not downwards-closed,
it might be tempting to consider its downward closure.
The following example shows that downward closures
interact poorly with pre-spectrality.%
\begin{example}\label{ex:prespectral:downclosure}
    We provide examples where $X = {\downarrow} Y$
    and~$X$ is pre-spectral but not $Y$ (resp.~$X$ is not pre-spectral
    but~$Y$ is).  Another instance will be given
    in \cref{ex:related:downcycles}.
    
    Consider $X \defined \mathbb{N} \uplus \{ \infty \}$
    with the ordering $x \leq y$ if and only if $x = y$ or $y = \infty$.
    The set $Y \defined \{ \infty \}$ with the Alexandroff topology
    is finite and therefore pre-spectral.  However,
    ${ \downarrow } Y = X$, which is not pre-spectral.

    Conversely, let $X \defined \Mod(\upsigma)$,
    $Y \defined \ModF(\upsigma)$, and $\leq$ be the reverse
    ordering~$\supseteq_i$, so that $X={\downarrow}Y$.  Remark that the
    space $\LPS{X}{\Alex{\leq}}{\FO[\upsigma]}$ is pre-spectral thanks
    to~\cref{thm:prespectral:rhpt} and the `dual' \ltrsk\ (a
    first-order sentence~$\varphi$ is preserved under induced
    substructures if and only if it is equivalent to a universal
    sentence).  However,
    {\ltrsk} does not relativise to finite
    structures~\cite{tait1959counterexample}
    and $\mathsf{UFO}$ generates a basis of the topology of $Y$,
    hence using~\cref{thm:prespectral:rhpt}
    the space $\LPS{Y}{\Alex{\leq}}{\FO[\upsigma]}$
    is not a \lpps.
\end{example}
\end{sendappendix}
\begin{remark}\label{rk:prespectral:lpps}
  For each of the fragments~$\mathsf{F}$ and associated
  quasi-orderings~$\leq$ of \cref{table:classical:cpres},
  $\LPS{\ModF(\upsigma)}{\Alex{\leq}}{\FO[\upsigma]}=(\ModF(\upsigma),\langle\Alex{\leq}\cap\modset{\FO[\upsigma]}_{\ModF(\upsigma)}\rangle)$
  is a pre-spectral space.  Indeed, by \cref{rk:alex:compact}, any
  compact open $K$ from $\CompSat{\ModF(\upsigma)}$ is the upward
  closure $K={\uparrow}F$ of a finite set $F\subfin \ModF(\upsigma)$,
  thus $K=\modset{\bigvee_{A\in
  F}\psi_A^\mathsf{F}}_{\ModF(\upsigma)}$, which shows that
  $\CompSat{\ModF(\upsigma)}\subseteq\modset{\mathsf
  F}_{\ModF(\upsigma)}$.  As any $\psi\in\mathsf F$ has finitely many
  $\leq$-minimal models in $\ModF(\upsigma)$,
  $\CompSat{\ModF(\upsigma)}\supseteq\modset{\mathsf
  F}_{\ModF(\upsigma)}$, and since $\mathsf F$ defines a base,
  $\LPS{\ModF(\upsigma)}{\Alex{\leq}}{\FO[\upsigma]}$ is pre-spectral.
  However, by \cref{cor:prespectral:rhpt} and the non-preservation
  results
  of~\cite{tait1959counterexample,gurevitch84,ajtai1994datalog,ajtai87,stolboushkin95},
  $\LPS{\ModF(\upsigma)}{\Alex{\subseteq_i}}{\FO[\upsigma]}$,
  $\LPS{\ModF(\upsigma)}{\Alex{\subseteq}}{\FO[\upsigma]}$, and $\LPS{\ModF(\upsigma)}{\Alex{\twoheadleftarrow}}{\FO[\upsigma]}$ are
  not \lpps: the condition
  $\uptau \cap \mathcal{L} \subseteq \CompSat{X}$ is crucial in order
  to derive preservation results.
\end{remark}

Another way of reaching the topological definitions of this section is
to consider a folklore result employed in several proofs of
preservation theorems over classes of finite structures for
fragments~$\mathsf F$ of
$\EFO$~\cite{Rossman08,atserias2008preservation,atserias2006preservation}:
if~$X$ is downwards-closed for~$\leq$, a monotone sentence~$\varphi$
is equivalent to a sentence from~$\mathsf F$ if and only if it has
finitely many $\leq$-minimal models in~$X$ (up to $\leq$-equivalence).
By~\cref{rk:alex:compact}, this says that $\modset{\varphi}_X$ is
compact, while the folklore result itself is essentially using the
fact that~$\mathsf F$ defines a base.

    \section{Related Notions}
    \label{sec:related}
    Pre-spectral spaces generalise two notions arising from order theory,
topology, and logics: Noetherian spaces and spectral spaces.

\subsection{Well-Quasi-Orderings and Noetherian Spaces}

A topological space in which all subsets are compact, or,
equivalently, all open subsets are compact, is called
\emph{Noetherian}~\cite[see][Section~9.7]{goubault2013non}.  A Noetherian
space $(X, \uptau)$ and a bounded sublattice $\mathcal{L}$ of $\parts
X$ always define a {\lpps} $\LPS{X}{\uptau}{\mathcal{L}}$.  A related
notion, considering a quasi-order instead of a topology, leads to the
well-known notion of
\emph{well-quasi-orders}~\cite{kruskal1972theory}: a quasi-order is a
well-quasi-order if and only if its Alexandroff topology is
Noetherian~\cite[Proposition~9.7.17]{goubault2013non}.  Thus, if
$(X,{\leq})$ is a well-quasi-order and $\mathcal{L}$ is a bounded
sublattice of $\parts X$, then $\LPS{X}{\Alex{\leq}}{\mathcal{L}}$ is
a \lpps.

\subparagraph{Applications of Noetherian Spaces to Preservation.}
Let us denote by~$\Graphs$ the class of finite
simple undirected graphs and by~$\sigmaG$ the signature with a single
binary edge relation~$E$; then the induced substructure
ordering~$\subseteq_i$ coincides with the induced subgraph ordering
over~$\Graphs$.

\begin{example}[Finite graphs of bounded tree-depth]
    Recall that the \emph{tree-depth} $\td(G)$ of a graph~$G$ is the
    minimum height of the comparability graphs~$F$ of partial orders
    such that~$G$ is a subgraph of~$F$~\cite[Chapter~6]{Neetil12}.
    Let $\Td_{\leq n}$ be the set of finite graphs of tree-depth at
    most~$n$ ordered by the induced substructure
    relation~$\subseteq_i$. This is a well-quasi-order~\cite{Ding92},
    thus $\langle\Td_{\leq
      n},\Alex{\subseteq_i},\FO[\sigmaG]\rangle$ is a
    \lpps, and therefore $\Td_{\leq n}$ enjoys the
    $(\Alex{\subseteq_i},\EFO[\sigmaG])$-preservation property
    by~\cref{thm:prespectral:rhpt}.
\end{example}

\begin{example}[Finite cycles]
    \label{ex:related:cycles}
    Consider the class $\Cycles\subseteq\Graphs$ of all finite simple
    cycles.  As is well known, $(\Cycles,\subseteq_i)$ is not a
    well-quasi-order because any two different cycles are incomparable
    for the induced substructure ordering~\cite{Ding92}.  In particular,
    every singleton is an open set: $(\Cycles, \Alex{\subseteq_i})$ is
    actually a topological space with the \emph{discrete topology},
    and its only compact sets are the finite sets:
    $\LPS{\Cycles}{\Alex{\subseteq_i}}{\FO[\sigmaG]}$ is not a
    {\lpps}.

    By standard locality arguments, for any sentence~$\varphi$, there
    exists a finite threshold~$n_0$ on the size of cycles, above
    which~$\varphi$ is either always true or always
    false\ifsubmission\ (see \appref{claim:related:cyclelocality})\fi.    
    \begin{sendappendix}
    
        \begin{ifappendix}
            \subparagraph{Finite Cycles (\cref{ex:related:cycles}).}
            We prove~\cref{claim:related:cyclelocality}
            stating that a formula~$\varphi$ cannot distinguish
            cycles above a certain size. Recall that~$\Cycles$
            is the set of cycles and that for~$n \geq 3$,
            $C_n$ is the simple cycle of size~$n$.
        \end{ifappendix}
    \begin{restatable}{claim}{claimrelatedcyclelocality}
    \label{claim:related:cyclelocality}
            For all $\varphi$ in~$\FO[\sigmaG]$, there exists a
            threshold~$n_0$ such that, for all $m,n\geq n_0$,
            $C_n \models \varphi$ if and only if $C_m \models \varphi$.
    \end{restatable}
        \begin{claimproof}
            Fix $d$ and $r$ two positive integers,
            and observe that for $n_0 \defined \max (2d+2, r+1)$,
            $n,m \geq n_0$, and $t$ an isomorphism type on~$d$-neighbourhoods
            of cycles,
            $C_m$ and $C_n$ have either both zero occurrences of~$t$ or both have more than~$r$ occurrences of~$t$.
            Thus, $n_0$ is
            such that for all $n,m \geq n_0$,
            $C_m$ and $C_n$ are \emph{$(r,d)$ threshold equivalent}~\cite[Definition~4.23]{libkin2012elements}.
            Fix a sentence $\varphi \in \FO[\sigmaG]$.
            By~\cite[Theorem~4.24]{libkin2012elements},
            there exists $(r,d)$ such that
            two $(r,d)$ threshold equivalent structures cannot be distinguished
            by~$\varphi$.
        \end{claimproof}

    We can directly show that~$\Cycles$ has the
    $(\Alex{\subseteq_i},\EFO[\sigmaG])$ preservation property
    using~\cref{claim:related:cyclelocality}.  For each~$n$, let
    $\psi_{\supseteq C_n}\in\EFO[\sigmaG]$ be the sentence defining the set of
    all structures that contain a cycle of size~$n$
    \begin{equation}
      \psi_{\supseteq C_n}\defined\exists
    x_0,\dots,x_{n-1}.\bigwedge_{i<j}\neg(x_i=x_j)\wedge\bigwedge_{0\leq
      i\leq n-1}E(x_i,x_{i+1\,\mathrm{mod}\,n-1})
    \end{equation}
    (thus $\modset{\psi_{\supseteq C_n}}_\Cycles=\{C_n\}$) and let
    $\psi_{\supseteq P_n}\in\EFO[\sigmaG]$ be the sentence defining
    the set of all structures that contain a path of size~$n$
    \begin{equation}
      \psi_{\supseteq P_n}\defined\exists
    x_0,\dots,x_{n-1}.\bigwedge_{i<j}\neg(x_i=x_j)\wedge\bigwedge_{0\leq
      i< n-1}E(x_i,x_{i+1})\;
    \end{equation}
    (thus $\modset{\psi_{\supseteq P_n}}_\Cycles=\{C_i\mid i\geq
    n\}$).
    
    If $\varphi$ is a monotone sentence, by
    \cref{claim:related:cyclelocality} either it has finitely many
    cycles $\{C_{n_1},\dots,C_{n_m}\}$ as models and $\bigvee_{1\leq
      j\leq N}\psi_{\supseteq C_j}$ is equivalent to it, or it has
    finitely many cycles $\{C_{n_1},\dots,C_{n_m}\}$ as counter-models
    and $\psi_{\supseteq P_{n_m+1}}\vee\bigvee_{\substack{j\leq
        n_m\\j\not\in\{n_1,\dots,n_m\}}}\psi_{\supseteq C_j}$ fits.

    \bigskip
    It is nevertheless enlightening to show preservation using the
    framework of logically defined pre-spectral spaces, by defining a
    suitable topology. %
    \end{sendappendix}
    Let $\uptau_n$ be the topology over~$\Cycles$
    generated by the definable
    co-finite sets and the definable sets containing only cycles of
    size at most~$n$.  This is a variation of the \emph{co-finite topology},
    and is also Noetherian\ifsubmission\ (see \appref{claim:related:cofinite})\fi.
    \begin{sendappendix}%
        \begin{ifappendix}%
            In~\cref{ex:related:cycles}, we
            used a variation the co-finite topology
            and claimed that it lead to a Noetherian space.
            We prove that claim below.
        \end{ifappendix}
    \begin{restatable}{claim}{claimrelatedcofinite}\label{claim:related:cofinite}
        The spaces $(\Cycles, \uptau_n)$ are 
        Noetherian for all~$n \in \mathbb{N}$
    \end{restatable}
        \begin{claimproof}
            Let $X \subseteq \Cycles$, and
            using Alexander's Subbase Lemma
            consider an open cover~$(U_i)_{i \in I}$ of~$X$ where
            $U_i$ either contains only cycles of size at most~$n$
            or is co-finite.
            
            If there exists~$i_0 \in I$ such that $U_i$ is co-finite,
            consider $X \setminus U_{i_0}$.  This is a finite set,
            hence there exists $I_0 \subfin I$ such that
            $X \setminus U_{i_0} \subseteq \bigcup_{i \in I_0} U_i$,
            which proves that $(U_i)_{i \in I_0 \cup \{ i_0 \}}$
            is a finite open cover of~$X$. If none of the~$U_i$
            is co-finite, then~$X$ is contained in the set of cycles
            of size at most $n$, which proves that~$X$ is finite,
            hence compact.            
        \end{claimproof}
    \end{sendappendix}
    Hence, $\langle\Td_{\leq
      n},\Alex{\subseteq_i},\FO[\sigmaG]\rangle$ is a \lpps, and
    as $\EFO[\sigmaG]$ defines a diagram base of
    it, we can apply
    \cref{thm:prespectral:rhpt} to deduce preservation.
    Now, given a monotone sentence~$\varphi$, either $\varphi$ has finitely many models or
    it has co-finitely many.  In both cases, this sentence defines an
    open set in~$\uptau_n$ for some~$n$ that is definable in $\EFO[\sigmaG]$.
    Thus the set of finite cycles has
    the $(\Alex{\subseteq_i},\EFO[\sigmaG])$ preservation property.
\end{example}
The previous example shows that the closure condition of
\cref{cor:prespectral:rhpt} was necessary, by proving that a space of
structures can enjoy a preservation theorem while not defining a
\lpps.
\begin{sendappendix}
  \Cref{ex:related:cycles} also shows that
  the hypothesis that $\mathsf F$ defines compact sets in
  \cref{thm:prespectral:rhpt}.\ref{gen-pres1} was necessary. Observe
  that the condition is violated by $\mathsf F=\EFO[\sigmaG]$
  in~$\Cycles$.  Indeed, $\modset{\psi_{\supseteq
      P_3}}_\Cycles\subseteq \bigcup_{n\geq 3}\modset{\psi_{\supseteq C_n}}_\Cycles$, but
  there does not exist a finite subcover of it.

  \begin{example}[Downward closure of cycles]\label{ex:related:downcycles}
    If one considers $\LPS{{\downarrow}
    \Cycles}{\Alex{\subseteq_i}}{\FO[\sigmaG]}$ the space of all the finite
    graphs that are induced substructures of some finite cycle, then
    \cref{cor:prespectral:rhpt} can be applied: there is preservation
    if and only if the space is a \lpps.  However that space
    is not a \lpps: consider the sentence~$\varphi$ stating that there
    exists no vertex of degree exactly~one; then $\modset{\varphi}_{{\downarrow}
    \Cycles}=\Cycles$, which is not compact, but is upwards-closed
    inside ${\downarrow}\Cycles$.  In particular, although 
    $\Cycles$ enjoys a preservation theorem,
    ${\downarrow} \Cycles$ does not.
  \end{example}
\end{sendappendix}

\subparagraph{Relativisation.}
The following proposition shows that, if we are looking for classes of
structures where preservation theorems \emph{always} relativize, then we
should endow them with a Noetherian topology.

\begin{proposition}
    Let $(X,\uptau)$ be a pre-spectral space
    such that for all $Y \subseteq X$, $Y$ with the induced
    topology is pre-spectral.  Then $X$ is Noetherian.
\end{proposition}
\begin{proof}
    Consider any subset $Y$ of $X$: by assumption, $Y$ is
    pre-spectral, hence compact in the induced topology, hence compact
    in~$(X,\uptau)$.
\end{proof}

    \subsection{Spectral Spaces}
\label{sec:spectral}

Spectral spaces are a class of topological spaces appearing naturally
in the study of logics and algebra as a generalisation of the Stone
Duality theory. Throughout this section we refer to two books and keep
the notations consistent with
them~\cite{goubault2013non,spectral2019}. A closed subset~$F$ of a
topological space~$X$ is \emph{irreducible} whenever~$F$ is non-empty
and is not the disjoint union of two non-empty closed sets.
The
\emph{closure} of a set~$Y$ in a space~$X$ is the smallest closed set
containing~$Y$ and is denoted by $\overline{Y}^X$
or $\overline{Y}$ when $X$ is clear from the context.
A topological
space~$X$ is \emph{sober} whenever any irreducible closed subset~$F$
is the closure of exactly one point~$x \in X$, which translates
formally to $\exists x \in X, \overline{\{x\}} = F$ and $\forall y \in
X, \overline{\{y\}} = F \Rightarrow y = x$.  A \emph{spectral space}
is a pre-spectral space that is
sober~\cite[Definition~1.1.5]{spectral2019}.

When a space $(X,\uptau)$ is not sober, it is possible to build a
\emph{sobrified} version of this space as
follows~\cite[Definition~8.2.17]{goubault2013non}: $\Sobrif{X}$ is the
set of irreducible closed sets of~$X$, and the topology is generated
by the sets $\Diamond U\defined \{ F \in \Sobrif{X} \mid F \cap U \neq
\emptyset \}$ where~$U$ is an open set of~$X$.  It can be shown that
this construction leads to a sober space, is idempotent up to
homeomorphism, and constructs the \emph{free} sober space over
$X$~\cite[Theorem 8.2.44]{goubault2013non}.  This leads to the
following correspondence between pre-spectral spaces and spectral
spaces.

\begin{fact}[Spectral versus pre-spectral]
    \label{lem:spectral:spectral}
    A space $X$ is pre-spectral if and only if
    $\Sobrif{X}$ is spectral.
\end{fact}

The connection with spectral spaces is of particular interest, because
the sobrification functor gives a tool to translate result
from the rich theory of spectral spaces to pre-spectral spaces
which will be extensively used in~\cref{sec:closure}.
\begin{sendappendix}For instance,
spectral spaces are closed under arbitrary products and
co-products~\cite[Theorem~2.2.1 and Corollary~5.2.9]{spectral2019}, and the sobrification operator commutes with these
operations~\cite[Fact~8.4.3 and Theorem~8.4.8]{goubault2013non}.
This allows to instantly deduce that a finite
disjoint union of pre-spectral
spaces is pre-spectral and an arbitrary product of pre-spectral spaces is
pre-spectral (see~\cref{lem:closure:unionps,lem:closure:prodps}).
\end{sendappendix}

\begin{sendappendix}
  \subparagraph{Spectral Spaces of Structures.}
  Spaces of finite structures, such as~$\Cycles$ the set of finite
  cycles, are in general not sober, and the spaces obtained through
  the constructions $\LPS{\Mod(\upsigma)}{\uptau}{\FO[\upsigma]}$ are
  not even~$T_0$. However, notice that the $T_0$ quotient of
  $\Mod(\upsigma)$ is sober when $\uptau = \Alex{\subseteq_i}$.  This
  proof can be adapted to the case where $\uptau = \Alex{\leq}$ and
  the upwards closure of finite structures are definable in
  $\Mod(\upsigma)$.  For a structure $A \in \Mod(\upsigma)$, define
  the \emph{age} of $A$ as $\Age(A) \defined \setof{A_0 \in
    \ModF(\upsigma)}{A_0 \subseteq_i A}$.
    \begin{claim}
        The $T_0$ quotient of
        $\LPS{\Mod(\upsigma)}{\Alex{\subseteq_i}}{\FO[\upsigma]}$
        is a sober space, hence a spectral space.
    \end{claim}
    \begin{claimproof}
        Remark that over $\ModF(\upsigma)$ the topology
        we consider is exactly $\Alex{\subseteq_i}$ since
        the upwards closure of a single finite structure
        is definable in $\FO[\upsigma]$.
        Notice moreover, that the topology over $\Mod(\upsigma)$
        is exactly the topology generated using sentences
        in $\EFO[\upsigma]$ thanks to the \ltrsk.
        In particular the $T_0$ quotient of $\Mod(\upsigma)$
        is only equating \emph{infinite} structures, 
        and two infinite structure are equated if and only if
        they have the same age.

        For all $A \in \ModF(\upsigma)$, let $\psi_A^\EFO$ be the
        diagram sentence such that $\modset{\psi^\EFO_A}_{\Mod(\upsigma)}$
        equals the upwards closure of $\{A\}$ in $\Mod(\upsigma)$.

        Consider $F$ an irreducible closed subset of $\Mod(\upsigma)$.
        Notice that for all $U,V$ open sets
        if $U \cap F \neq \emptyset$ and $V \cap F \neq \emptyset$
        then $U \cap V \cap F \neq \emptyset$.
        Assume by contradiction that a pair $U,V$ exists
        such that $U \cap F \neq \emptyset$ and $V \cap F \neq \emptyset$,
        but $U \cap V \cap F = \emptyset$. Then
        $F \subseteq U^c \cup V^c$, but $F \subsetneq U^c$
        and $F \subsetneq V^c$. Since $F$ is closed,
        we can write $F = (U^c \cap F) \cup (V^c \cap F)$
        an conclude that $F$ is the disjoint union of two non-empty
        closed sets which is absurd because $F$ is irreducible.

        Because $F$ is closed in $\Mod(\upsigma)$ it is
        a downwards closed set for $\subseteq_i$.
        Assume that $A$ and $B$ are two finite structures
        in $F$, remark that
        $F \cap \modset{\psi_A^\EFO}_{\Mod(\upsigma)} \neq
        \emptyset$
        and 
        $F \cap \modset{\psi_B^\EFO}_{\Mod(\upsigma)} \neq
        \emptyset$; since $F$ is irreducible, this proves
        the existence of $C \in F$ such that
        $C \in F \cap \modset{\psi_A^\EFO \wedge \psi_B^\EFO}_{\Mod(\upsigma)}$.
        In particular, $A \subseteq_i C$ and $B \subseteq_i C$ and $C \in F$.
        Note that $C$ might not be finite, but the result can be extended
        to a finite set of structures $A_1, \dots, A_n$:
        if $A_1, \dots, A_n \in F \cap \ModF(\upsigma)$, then
        there exists $C$ above all of them that is still in $F$.

        Let $T^+ \defined \setof{\psi_A^\EFO}{A \in F \cap \ModF(\upsigma)}$
        and $T^- \defined \setof{\neg \psi_A^\EFO}{A \in F^c \cap
        \ModF(\upsigma)}$.
        Remark that any
        finite subset of $T^+$ has a model in $F$, this is because
        a finite subset of $T^+$ defines the upwards closure of a finite
        number of finite structures in $F$, hence has a (possibly infinite)
        model in $F$.
        Remark that an element of $F$ is a model of $T^-$,
        because $F$ is downwards closed for $\subseteq_i$.
        By the
        Compactness Theorem, $T^+ \cup T^-$ has a model $B$.
        The definition of $T^+ \cup T^-$ immediately
        ensures that $\Age(B) = F \cap \ModF(\upsigma)$.

        Let us show that $B \in F$.
        Assume by contradiciton that $B \in F^c$, then there exists
        an open set $U$ defined by $\varphi \in \EFO[\upsigma]$
        such that $B \models \varphi$ and $U \subseteq F^c$.
        As $\varphi \in \EFO[\upsigma]$, it has a finite model $A_0$
        such that $A_0 \subseteq_i B$.
        This is absurd because it shows that $A_0 \not \in F$,
        $A_0 \in \ModF(\upsigma)$
        but $A_0 \in \Age(B) = F \cap \ModF(\upsigma)$.

        Let $A \in F \cap \ModF(\upsigma)$,
        let us show that $A$ is in the closure of $\{B\}$.
        For this, we consider a sentence $\varphi \in \EFO[\upsigma]$
        such that $A \models \varphi$, and show that $B \models \varphi$.
        Given such a sentence $\varphi$,
        remark that $\modset{\psi_{A}^\EFO}_{\Mod(\upsigma)}
        \subseteq \modset{\varphi}_{\Mod(\upsigma)}$
        because $\varphi$ is monotone.
        By construction, $B \models \psi_{A}^\EFO$,
        hence $B \models \varphi$.
        Hence, $F \cap \ModF(\upsigma)$ is included in the closure of $B$.

        Now, consider
        an infinite structure $B' \in F$,
        we are going to prove that $B'$ is in the closure of $\{B\}$.
        For that, consider $\varphi \in \EFO[\upsigma]$
        such that $B' \models \varphi$, as $\varphi \in \EFO[\upsigma]$
        there exists a finite structure $A \subseteq_i B'$ such that
        $A \models \varphi$. Because $F$ is downwards-closed,
        $A \in F$, hence $A$ is in the closure of $\{B\}$, therefore $B \models
        \varphi$.

        As $F$ is a closed set and $B \in F$,
        the closure of $B$ is included in $F$ and we have
        proven that $F$ is the closure of a point in $\Mod(\upsigma)$.
        In the $T_0$ quotient of $\Mod(\upsigma)$ this point
        is unique by definition.

        We have proven that every non-empty irreducible closed
        set is the closure of exactly one point in the $T_0$
        quotient of $\Mod(\upsigma)$, which is the definition
        of a sober space.
    \end{claimproof}
\end{sendappendix}

    \section{Basic Closure Properties}
    \label{sec:closure}
    To study preservation theorems, we not only want to ensure that the
space is pre-spectral, but also to see that the lattice of compact
open sets is obtained through a restriction of the logic.  Therefore,
one of our main concerns with closure properties is to characterise
the lattice of compact sets, which must use properties of the
definable sets and cannot rely solely on topological constructions.

\subsection{Morphisms}

\subparagraph{Spectral Maps.}
Let us first introduce the notion of morphism between pre-spectral
spaces, inherited from the case of spectral spaces~\cite[Definition
1.2.2]{spectral2019}.  A map $f \colon (X,\uptau) \to (Y, \uptheta)$
is a \emph{spectral map} whenever it is continuous and the pre-image
of a compact-open set of $Y$ is a compact-open set of $X$. We will
write~$\CatPS$ for the category of pre-spectral spaces and spectral
maps.

\begin{fact}
    The image of a pre-spectral space through a
    spectral map is pre-spectral.
\end{fact}
A crucial role of spectral maps is to guard the definition
of \emph{pre-spectral subspaces}, mimicking the one of \emph{spectral
subspaces}~\cite[Section~2.1]{spectral2019}. A pre-spectral subspace
is not only a subset where the induced topology happens to be
pre-spectral, but has the additional property that the \emph{inclusion
map} is a spectral map.

\subparagraph{Logical Maps.}
In the case of a \lpps, a map $f
\colon \LPS{X}{\uptau}{\mathcal{L}} \to \LPS{Y}{\uptheta}{\mathcal{L}'}$ is
a \emph{logical map} whenever it is continuous and the pre-image of a
definable open set of $Y$ is a definable open set of $X$.  A map
between logically defined pre-spectral spaces is logical if and
only if it is spectral, since compact open subsets and definable
open subsets coincide in that case.  However, the use of logical maps
is to prove that some spaces are pre-spectral by transferring logical
properties rather than topological ones.
\begin{fact}
    The image of a {\lpps}
    $\LPS{X}{\uptau}{\mathcal{L}}$ through a logical map is
    a \lpps.
\end{fact}

Of particular interest are the logical maps obtained through syntactic
constructions.  Let us define an \emph{$\FO$-interpretation} $f \colon
X \to Y$ where $X \subseteq \Mod(\upsigma_1)$ and
$Y \subseteq \Mod(\upsigma_2)$ through `relation' formul\ae\ $\rho_R$
for all $R \in \upsigma_2$, where $\rho_R$ has as many free variables
as the arity of $R$, and an additional `domain' formula
$\delta \in \FO[\upsigma_1]$ with one free variable.  The image of a
$\upsigma_1$-structure $A \in X$ is the $\upsigma_2$-structure $f(A)$
with domain $|f(A)| \defined \setof{a \in |A|}{A\models \delta(a) }$
and such that $(a_1, \dots, a_n)\in\Rel{R}{f(A)} \iff
A\models \rho_R(a_1, \dots, a_n)$.  This is a simple model of logical
interpretations: many different notions %
can be found in the literature~\cite[see][]{courcelle94}.

An $\FO$-interpretation $f \colon X \to Y$ allows to transfer logical
properties from one class of structures to another: if
$\varphi \in \FO[\upsigma_2]$ is a formula on the structures of~$Y$,
then there exists a formula~$f^{-1}(\varphi) \in \FO[\upsigma_1]$ such
that $A\models f^{-1}(\varphi)(\vec{a})$ if and only if
$f(A)\models \varphi(f(\vec{a}))$~\cite[Section
4.3]{hodges1997shorter}; thus, the pre-image of a definable
set is definable.

\begin{fact}
    An $\FO$-interpretation
    is a logical map if and only if it is continuous.
\end{fact}

This provides us with a proof scheme to show that a space
$\LPS{Y}{\uptau_2}{\FO[\upsigma_2]}$ is a \lpps: first, build a
{\lpps} $\LPS{X}{\uptau_1}{\FO[\upsigma_1]}$, then build a
$\FO$-interpetation that is surjective and continuous from~$X$ to~$Y$,
and conclude that $Y$ is a \lpps. This is used for instance by
\citet[Corollary~10.7]{Neetil12} to show that the class of all
$p$~\emph{subdivisions} of finite graphs enjoys homomorphism
preservation (using a slightly more general notion of
$\FO$-interpretations).

\subsection{Relativisation}

Preservation theorems do not
relativise in general, but the stronger notion of being pre-spectral
shows that non trivial sufficient conditions for relativisation
exists.  However, unlike the theory of spectral spaces, there is not
yet a full characterisation of the pre-spectral subsets of a
pre-spectral space\ifsubmission; see \appref{app:sec:closure} for a
discussion\fi.

\begin{sendappendix}
  \begin{ifappendix}
    \subparagraph{Pre-spectral Subspaces.}
    Recall that $(Y,\uptheta)$ is a \emph{pre-spectral subspace} of $(X,\uptau)$
    whenever $(Y,\uptheta)$ is a pre-spectral space such that $Y \subseteq X$, 
    $\uptheta$ is the topology induced by $\uptau$ on $Y$,
    and the inclusion map is spectral.
  \end{ifappendix}
  \begin{restatable}[Sufficient condition for pre-spectral
    subspaces]{lemma}{lemclosuresubset}\label{lem:closure:subset}
    Every positive Boolean combination $X$ of closed subsets of $Y$
    and compact-open subsets of $Y$ defines a pre-spectral subspace of $Y$.
  \end{restatable}
  \begin{proof}
    It suffices to prove that for all compact-open set $U$ of $X$
    the set $U \cap Y$ is a compact-open set of $X$.
    Note that in particular this proves
    $\CompSat{Y} \supseteq \setof{ U \cap Y}{U \in
    \CompSat{X}}$.
    Note that this stronger property is preserved
    under finite unions and finite intersections. Therefore,
    we only consider the case of a closed subset, or a compact-open subset.
    When $F$ is a closed subset of $X$, we use the fact that the intersection
    of a compact and a closed subset defines a compact subset.
    When $F$ is a compact open set of $X$, we use the fact that $\CompSat{X}$
    is a lattice to conclude.
  \end{proof}

  \begin{example}[Pre-spectral vs.\ logically presented
    pre-spectral]\label{ex:closure:psvslpps}
    An interesting example that will be further discussed
    in \cref{ex:spectral:nosubset} is $\mathcal{D}_{\leq
    2}\subseteq\Graphs$ the set of finite simple graphs of degree
    bounded by two.  It turns out that 
    $\LPS{\mathcal{D}_{\leq 2}}{\Alex{\subseteq_i}}{\FO[\upsigma]}$ is
    a \lpps\ by~\cref{cor:prespectral:rhpt} and the fact that
    $\mathcal{D}_{\leq 2}$ is downwards-closed and has the
    $(\Alex{\subseteq_i},\EFO[\upsigma])$ preservation
    property~\cite{atserias2008preservation}.  As seen
    in \cref{ex:related:downcycles}, the space of downward closures of
    finite cycles 
    $\LPS{{\downarrow} \Cycles}{\Alex{\subseteq_i}}{\FO[\upsigma]}$ is
    not a \lpps.
        
    However, observe that $Y\defined{\downarrow} \Cycles$ is a closed
    subset of $X\defined\mathcal{D}_{\leq 2}$, because
    $\mathcal{D}_{\leq 2} \setminus {\downarrow} \Cycles$ is the set
    of graphs containing the disjoint union of a cycle and an isolated
    vertex as a subgraph.  As we just saw
    in \cref{lem:closure:subset}, a closed subset of a pre-spectral
    space is pre-spectral, thus $({\downarrow} \Cycles, \uptau_Y)$
    where $\uptau_Y$ is the topology induced by~${\downarrow} \Cycles$
    on $\langle\modset{\FO[\upsigma]}_{\mathcal{D}_{\leq
    2}} \cap \Alex{\subseteq_i}\rangle$ is a pre-spectral space,
    albeit not a logically presented one.

    The reason behind this apparent discrepancy is that, when restricting
    the set of structures, more sentences become monotone. 
In particular, considering $Y'$ as $Y$ with the induced topology from
    $X$ we have $\CompSat{Y} = \CompSat{Y'}
    = \{ \modset{\varphi}_{Y'} \mid \modset{\varphi}_X \in \Alex{\subseteq_i} \} \subsetneq \modset{\FO[\upsigma]}_{Y} \cap \Alex{\subseteq_i}$;
    see \cref{fig:closure:differentsubspaces} for an illustration.
    \begin{figure}
        \centering
\def\svgwidth{0.6\columnwidth}
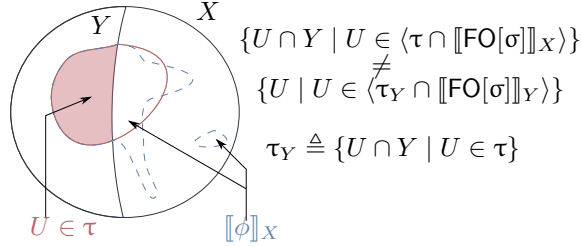

        \caption{Different induced subspace constructions.}
        \label{fig:closure:differentsubspaces}
    \end{figure}
  \end{example}
\end{sendappendix}

\begin{proposition}[Sufficient condition for relativisation]
    \label{thm:closure:subspace}
    Let $\LPS{X}{\uptau}{\FO[\upsigma]}$ be a \lpps,
    $Y$ be a Boolean combination of compact-open subsets of~$X$,
    and $\uptheta$ be the topology induced by~$\uptau$ on~$Y$.
    Then $\LPS{Y}{\uptheta}{\FO[\upsigma]}$ is a \lpps.
\end{proposition}
\begin{proof}
    It suffices to prove that any definable open set~$U$ of~$Y$
    is the restriction of some definable open set of~$X$ to~$Y$.
    This stronger hypothesis is stable under finite unions
    and finite intersections, thus we only need to deal with the cases
    where~$Y$ is a definable open of~$X$ or the complement of one.

    Let us first consider the case where
    $Y$ is a definable open set %
    of~$X$.  %
    Then $U = U \cap Y$ is the restriction of an open definable set
    of~$X$ to~$Y$.  Let us next consider the case where~$Y$ is a
    definable closed set of~$X$.  Remark that $V \defined U \cup (X \setminus
    Y)$ is an open set  of~$X$, and is still definable. Therefore $U =
    V \cap Y$ with~$V$ a definable open set of~$X$.
\end{proof}

\begin{sendappendix}
\subparagraph{Pro-constructible Sets.}
One hope could be more generally to translate the characterisation of
spectral subspaces of a spectral space to pre-spectral spaces.  In a
spectral space, a spectral subspace is a \emph{pro-constructible set},
that is a set written as $\bigcap_{i \in I} (U_i^c \cup V_i)$ where
$U$ and $V$ are compact open sets~\cite[Definition~1.3.10,
Proposition~1.3.13, Theorem~2.1.3]{spectral2019}.  We will use the
notation $U
\Rightarrow V$ instead of $U^c \cup V$ to understand these sets as
satisfying a theory of Horn clauses.

In a {\lpps} $\LPS{X}{\uptau}{\FO[\upsigma]}$, pro-constructible sets
correspond precisely to the models of a \emph{first-order theory}
of sentences in $\FO[\upsigma]$ that are open in $\uptau$ or
closed in $\uptau$.
Pro-constructible subsets of $\LPS{X}{\uptau}{\FO[\upsigma]}$
are \lpps, for any topology $\uptau$ on the set of structures.
This is a consequence of the fact that a pro-constructible set
is in particular defined by a first-order theory,
hence the Compactness Theorem of first-order logic still applies
to it.
However, this relativisation does not generalise,
as demonstrated by~\cref{ex:spectral:nosubset} below.

\begin{example}[Pro-constructible sets are not sufficient]
    \label{ex:spectral:nosubset}Consider again the spaces
    from~\cref{ex:closure:psvslpps}. The subset $\Cycles \subseteq \mathcal{D}_{\leq 2}$
    consisting of all the finite cycles is definable using a
    \emph{pro-constructible set}.  For instance, one can write that it
    contains no graph containing a disjoint union of a cycle and a
    point, and add a constraint that any graph that is not above a
    cycle of size~$n$ contains at least~$n+1$ points.  Moreover, the
    upward closure of a cycle is an open set in~$\mathcal{D}_{\leq
      2}$, but is a singleton when restricted to $\Cycles$, proving
    that the space is not pre-spectral because it has the discrete
    topology and is infinite.
\end{example}

This counter-example shows that one cannot solely rely on the
properties of spectral spaces, but actually have to work to translate
their results.
Moreover, this example also provides a {\lpps} that is~$T_0$
but not sober.
\Cref{ex:spectral:nosubset} proves that $\mathcal{D}_{\leq 2}$
is a \lpps, and that it has a pro-constructible
subset that is not pre-spectral (hence not spectral).
However, if $\mathcal{D}_{\leq 2}$ were sober, it would be
spectral, and any pro-constructible subset would be
spectral, which is a contradiction.

\Cref{ex:spectral:nosubset} also shows that we cannot strengthen
the hypothesis of~\cref{lem:closure:subset} to consider arbitrary intersections.
One might wonder whether the characterisation of spectral subspaces
as sets of the form $\bigcap_{i \in I} U_i \Rightarrow V_i$ where
$U_i$ and $V_i$ are compact open sets could lead to a necessary condition
on pre-spectral spaces.
The following example answers negatively.

\begin{example}[Pro-constructible sets are not necessary]
    Consider $X \defined \mathbb{N}$ with the Alexandroff topology
    over its natural order $\leq$.
    Consider $Y \defined \Sobrif{X}$,
    that is $Y \simeq \mathbb{N} \cup \{ \infty \}$ with
    the Scott topology of the natural order over $Y$.
    Both $X$ and $Y$ are Noetherian spaces, hence pre-spectral.
    Moreover, $Y$ is sober, hence $Y$ is a spectral space.
    The inclusion of $X$ into $Y$ is spectral because
    $X$ is Noetherian.
    Hence $X$ is a pre-spectral subspace of $Y$.
    However, notice that if $X$ was obtained as a (possibly infinite)
    intersection of finite Boolean combinations of
    compact-open sets of $Y$,
    $X$ would be sober~\cite[Corollary~3.5]{keimel2009292}.
    However, $X \neq Y$ and $\Sobrif{X} = Y$, which is a contradiction.
\end{example}

\end{sendappendix}

\subsection{Disjoint Unions and Products}

Rather than using an already existing pre-spectral space and
considering sub-spaces to build new smaller ones,
it can be a rather efficient method to combine existing spaces
to build bigger spaces.
However, to build preservation theorems out of these constructions,
it is necessary to represent those them
as spaces of structures over some relational signature,
which will be the role of~\cref{def:closure:sumspace,def:closure:prodspace}. 

\begin{sendappendix}
    \subparagraph{Finite Disjoint Unions.}
Spectral spaces are closed under arbitrary
co-products \cite[Corollary 5.2.9]{spectral2019}, but these
co-products are in general not obtained through simple
disjoint unions. Indeed, an infinite discrete space is not spectral
because it is not compact, but it can
be obtained as an arbitrary disjoint union of singletons
(which are spectral spaces).
The characterisation of co-products of spectral spaces
is the goal of an entire section in~\cite[Section~10.1]{spectral2019}.

Since we are interested in building new classes of structures, we
restrict ourselves to the cases where a concrete representation exists
in terms of classes of structures so that we can interpret first-order
logic on the co-product; therefore, we only study finite sums.
    The following lemma is a then direct
    application of the stability of spectral spaces under finite disjoint
    union.
\begin{lemma}[Stability under finite disjoint union]
    \label{lem:closure:unionps}
    Let $(X_i, \uptheta_i)_{i \in I}$ be a family of pre-spectral spaces,
    where the index set $I$ is finite.
    The space $\sum_{i \in I} X_i$ with the sum topology
    is a pre-spectral space.
\end{lemma}
\begin{proof}
    By \cref{lem:spectral:spectral}, $(\Sobrif{X_i}, \Sobrif{\uptheta_i})$
    is a spectral space.
    However, spectral spaces are stable under topological
    sums~\cite[Corollary 2.4.4]{spectral2019},
    and $\Sobrif{\sum_{i \in I} X_i} \simeq \sum_{i \in I} \Sobrif{X_i}$
    thanks to~\cite[Fact~8.4.3]{goubault2013non}.
    Therefore, by \cref{lem:spectral:spectral}, $\sum_{i \in I} X_i$ is a
    pre-spectral space.
\end{proof}
\end{sendappendix}

\begin{definition}[Logical sum]
    \label{def:closure:sumspace}
    Let $(\LPS{X_i}{\uptau_i}{\FO[\upsigma_i]})_{i \in I}$ be a family of
    spaces. The \emph{logical sum} $\LPS{X}{\uptau}{\FO[\upsigma]}$
    is defined as follows:
    \begin{enumerate}
        \item The signature $\upsigma$ is the disjoint union of the signatures
        $(\upsigma_i)_{i \in I}$.
        \item The set $X$ is the union (disjoint by construction)
 $\bigcup_{i \in I}
    f_i(X_i)$ where, for all $i \in I$, $f_i : X_i \to \Mod(\upsigma)$ is defined
        by $|f_i(A)|\defined |A|$ and $(a_1, \dots, a_n)\in\Rel{R}{f_i(A)}$ if and only if
        $R \in \upsigma_i$ and $(a_1, \dots, a_n)\in\Rel{R}{A}$.
        \item The topology $\uptau$ is generated by the
        sets $f_i (U)$ where $U \in \uptau_i$ and $i \in I$.
    \end{enumerate}
\end{definition}

The logical sum space is a simple translation of the topological sum
space, which leads to the following result\ifsubmission\
(see \appref{app:sec:closure})\else. In this sense, 
\cref{lem:closure:unionps} is mostly a rephrasing\fi.
\begin{restatable}[Stability under finite logical sum]{proposition}{lemclosureunionlps}
    \label{lem:closure:unionlps}
    Let $(\LPS{X_i}{\uptau_i}{\FO[\upsigma_i]})_{i \in I}$ be a finite
    family of \lpps.
    The logical sum of those spaces
    is a {\lpps} homeomorphic to the sum of those spaces in $\CatPS$.
\end{restatable}

\begin{sendappendix}
    \begin{ifappendix}
        \lemclosureunionlps*
    \end{ifappendix}

\begin{proof}
    Recall that if $\LPS{X_i}{\uptau_i}{\FO[\upsigma_i]}$
    is a {\lpps},
    then $(X_i, \uptheta_i)$ is a pre-spectral space,
    with $\uptheta_i \defined \langle \uptau_i \cap
        \modset{\FO[\upsigma_i]}_{X_i} \rangle$
    and $\CompSat{X_i} = 
        \uptau_i \cap \modset{\FO[\upsigma_i]}_{X_i}$.
    By~\cref{lem:closure:unionps}, the topological sum of these spaces $Y \defined
        \sum_{i \in I} (X_i, \uptheta_i)$ is a
    pre-spectral space.  Recall that $\langle
    Y,\uptau_Y,\CompSat{Y}\rangle$ is a \lpps. We are going to exhibit
    a surjective map $f$ from $Y$ to the logical sum~$X$ of
    the~$X_i$.

    To a structure $A \in X_i$, we associate $f(A) \defined f_i (A)$. By
    definition, this map is surjective. Notice that this map is actually
    a \emph{logical map} from $\LPS{Y}{\uptau_Y}{\CompSat{Y}}$ to
    $\LPS{X}{\uptau}{\FO[\upsigma]}$. Indeed, consider a definable open set of
    $X$: it is obtained through a sentence
    $\varphi \in \FO[\upsigma]$.  However,
    a simple rewriting allows us to write $\varphi \equiv_X \varphi_1 \vee \cdots\vee
        \varphi_k$ with $\varphi_k \in \FO[\upsigma_{i_k}]$ with $i_1, \dots, i_k \in I$.
    By definition of the sum topology, the sets $\varphi_{i_j}$ are therefore open in
    $X_{i_j}$ and thus compact because each
    $\LPS{X_{i_j}}{\uptau_{i_j}}{\FO[\upsigma_{i_j}]}$ is a \lpps.
    In particular, the pre-image of an open definable set of $X$ is a compact
    open set of~$Y$, proving that $f$~is continuous, and logical.
\end{proof}

  \subparagraph{Products.}%
\end{sendappendix}

In the case of products, a sentence over a product is not simply
obtained by projecting on each component.  This is handled in our
proof of \cref{lem:closure:prodlps}\ifsubmission\
in \appref{app:sec:closure}\fi\ by reducing the first-order theory of
the product to the first-order theories of its components thanks to
Feferman-Vaught
decompositions~\cite{feferman1959first,makowsky2004algorithmic}.

\begin{sendappendix}
   Again, we first state a direct application of the stability of
   spectral spaces under products.
\begin{lemma}[Stability under products]
    \label{lem:closure:prodps}
    Let $(X_i, \uptheta_i)_{i \in I}$ be a family of pre-spectral spaces.
    The product space $X \defined \prod_{i \in I} X_i$ with the product topology is
    pre-spectral.
    Moreover, compact open sets of $X$
    are obtained as finite union of
    $p_i^{-1} (K)$ with $i \in I$ and $K \in \CompSat{X_i}$
    where $p_i$ is the projection over the $i$th component.
\end{lemma}
\begin{proof}
    By \cref{lem:spectral:spectral}, $(\Sobrif{X_i}, \Sobrif{\uptheta_i})$
    is a spectral space. Since spectral spaces are stable under
    products~\cite[Theorem 2.2.1]{spectral2019}, and
    $\Sobrif{\prod_{i \in I} X_i} \simeq \prod_{i \in
    I} \Sobrif{X_i}$~\cite[Theorem~8.4.8]{goubault2013non}.
    Therefore, using \cref{lem:spectral:spectral}, $\prod_{i \in I} X_i$ is a
    pre-spectral space.

    Consider a compact open set $U$ of the product
    and $i \in I$, the image $p_i(U)$ is an open set of $X_i$
    by definition of the product topology, and is compact because $p_i$
    is continuous. Conversely, let $i \in I$ and
    let $U_i$ be a compact open set of $X_i$,
    define $V_j \defined X_j$ for $j \neq i \in I$, and $V_i \defined U_i$.
    Then $p_i^{-1}(U_i) = \prod_{i \in I} V_i$
    is compact as a product of compact spaces
    thanks to Tychonoff's Theorem. Moreover, it is an open set
    because $p_i$ is continuous.
\end{proof}
\end{sendappendix}

\begin{definition}[Logical product]
    \label{def:closure:prodspace}
    Let $(\LPS{X_i}{\uptau_i}{\FO[\upsigma_i]})_{i \in I}$ be a family of
    spaces. The \emph{logical product} $\LPS{X}{\uptau}{\FO[\upsigma]}$
    is defined as follows:
    \begin{enumerate}
        \item The signature $\upsigma$ is the disjoint union of the signatures
            $(\upsigma_i)_{i \in I}$
            with additional unary predicates $\varepsilon_i$ for
            each~$i\in I$.
        \item The set $X$ is the image of $\prod_{i \in I} X_i$
            through the map $f \colon \prod_{i \in I} X_i \to \Mod(\upsigma)$
            that associates to each $(A_i)_{i \in I}$
            the disjoint union of the structures~$A_i$
            with $\varepsilon_i$ true on the structure $A_i$
            for $i \in I$.
        \item The topology $\uptau$
            generated by the sets $U$
            such that $f^{-1}(U)$ is an open set
            of
            $\prod_{i \in I} (X_i, \uptau_i)$.
    \end{enumerate}
\end{definition}

\begin{restatable}[Stability under finite logical product]{proposition}{lemclosureprodlps}
    \label{lem:closure:prodlps}
    Let
    $(\LPS{X_i}{\uptau_i}{\FO[\upsigma_i]})_{i \in I}$
    be a finite family of \lpps.
    The logical product of those spaces
    is a {\lpps} homeomorphic
    to the product of the spaces $X_i$ in $\CatPS$.
\end{restatable}

\begin{sendappendix}
\begin{ifappendix}
    \lemclosureprodlps*
\end{ifappendix}
\begin{proof}
    We consider a binary product
    and by an immediate induction conclude that the theorem
    holds for finite products.
    Consider $X$ the logical product of two {\lpps}
    $\LPS{X_1}{\uptau_1}{\FO[\upsigma_1]}$ and
    $\LPS{X_2}{\uptau_2}{\FO[\upsigma_2]}$.
    We prove that the map
    $f \colon X_1 \times X_2 \to X$
    that associates to each $(A_1, A_2)$
    the disjoint union $A_1 \uplus A_2$
    is a logical map. This will allow us to conclude because
    $X_1 \times X_2$ is pre-spectral as a product of two
    pre-spectral spaces by~\cref{lem:closure:prodps}.

    \medskip
    To this end, we prove that 
    the pre-image of a non-empty definable open set of $X$
    is a compact open set of $X_1 \times X_2$.
    Consider such a definable open set $U$: by definition of the topology
    over the logical product, it is the image of
    an open set of $(X_1, \uptau_1) \times (X_1, \uptau_2)$.
    Because $U$ is definable,
    $U = \modset{\varphi}_X$ for some $\varphi \in \FO[\upsigma]$.
    By the Feferman-Vaught Decomposition Theorem
    over finite disjoint unions~\cite{feferman1959first},
    there exists finitely many sentences $(\psi_i^j)_{1 \leq i \leq
      n}^{j \in \{1,2\}}$ and a Boolean function $\beta \colon \{0,1\}^{2n} \to \{0,1\}$
            such that
            \begin{equation}\label{eq:fv}
                \forall x \in X_1, y \in X_2,
                f(x,y) \models \varphi 
                \iff \beta\left( (x \models \psi_i^1)_{1 \leq i
                \leq n} ; (y \models \psi_i^2)_{1 \leq i \leq n}\right) = 1\;.
            \end{equation}

            We build two equivalence relations of finite index 
            respectively on $X_1$ and $X_2$.
            The relation $\equiv_1$ over $X_1$ is defined
            by $x \equiv_1 x'$ if
            $\forall 1 \leq i \leq n,  x \models \psi_i^1 \iff
            x' \models \psi_i^1$.
            The relation $\equiv_2$ over $X_2$ is defined by
            $y \equiv_1 y'$ if
            $\forall 1 \leq i \leq n,  y \models \psi_i^2 \iff
            y' \models \psi_i^2$.
            We let $(x,y) \equiv (x',y')$ if and only if
            $x \equiv_1 x'$ and $y \equiv_2 y'$.
            Thus, by~\eqref{eq:fv}, if $f(x,y) \models \varphi$
            and $(x,y) \equiv (x',y')$, then $f(x',y') \models \varphi$.
            This proves that $f^{-1}(U)$ is a subset of $X_1 \times X_2$
            that is saturated for $\equiv$.

            For $i \in \{ 1, 2 \}$,
            a subset $U \subseteq X_i$ that is saturated for $\equiv_i$
            is definable in $\FO[\upsigma_i]$
            since it can be obtained as a finite Boolean combination
            of the sets $( \modset{\psi_j^i}_{X_i} )_{1 \leq j \leq
              n}$.  Thus, if $V_1 \in \uptau_1$, $V_2 \in \uptau_2$,
            and $V_1 \times V_2$ is saturated for~$\equiv$,
            then $V_1$ (resp.\ $V_2$) is definable in $X_1$ (resp.\ $X_2$).
            Hence using the fact that $X_1$
            and $X_2$ are \lpps,
            $V_1$ and $V_2$ are compact open sets of $X_1$ and $X_2$.
            By Tychonoff's Theorem the product $V_1 \times V_2$ is
            compact in $X_1 \times X_2$.
            We are going to prove that $f^{-1}(U)$ is actually
            a finite union of $\equiv$-saturated cylinder sets
            of this form
            and conclude using the fact that a finite union of compact
            open sets is a compact open set.

        \medskip
        First of all,
            because $f^{-1}(U)$ is open in $(X_1, \uptau_1) \times (X_2,
            \uptau_2)$, it is obtained as a union of sets
            of the form $V_1 \times V_2 \subseteq f^{-1}(U)$
            with $V_1 \in \uptau_1$ and $V_2 \in \uptau_2$
            and both are non-empty.
            Consider a set $V_1 \times V_2 \subseteq f^{-1}(U)$ with
            $V_1 \in \uptau_1$ and $V_2 \in \uptau_2$ that is
            \emph{maximal} for inclusion; such a set exists by Zorn's
            Lemma.  We are going to prove that $V_1 \times V_2$ is
            $\equiv$-saturated.

    \begin{figure}[tb]
        \centering
\def\svgwidth{0.8\columnwidth}
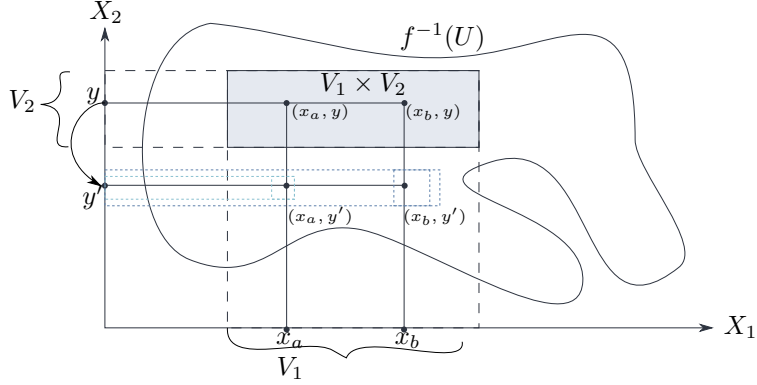

        \caption{Saturation of maximal cylinder open subsets in the case of a binary
        product $X \times Y$.}
        \label{fig:closure:prodcylinder}
    \end{figure}

        Assume by contradiction that $V_1 \times V_2$ is non empty
            and
            is not $\equiv$-saturated; without loss of generality,
            assume that $V_2$ is not $\equiv_2$-saturated.
            Following~\cref{fig:closure:prodcylinder},
            this shows that there exists $y \in V_2$
            and $y' \not \in V_2$ such that $y \equiv_2 y'$.
            Remark that for any $x \in V_1$,
            $(x,y) \equiv (x,y')$, hence $f(x,y) \in U$
            implies $f(x,y') \in U$.

        Because $f^{-1}(U)$ is open in $(X_1, \uptau_1) \times (X_2,
            \uptau_2)$, whenever $(x,y') \in f^{-1}(U)$,
            there exists a cylinder $V_1^{(x,y')} \times V_2^{(x,y')}$
            with $V_1^{(x,y')} \in \uptau_1$, $V_2^{(x,y')} \in \uptau_2$
            such that
            $(x,y') \in V_1^{(x,y')} \times V_2^{(x,y')} \subseteq f^{-1}(U)$.
            We illustrate this using two points $x_a$ and $x_b$
            in~\cref{fig:closure:prodcylinder}.
        Notice that $V_2^{(x,y')}$ is an open neighborhood of $y'$ for
        all $x \in V_1$.

        Because $\equiv_1$ is of finite index, there
        exists a finite set $F \subseteq V_1$ such that $\forall x \in
        V_1, \exists x' \in F, x \equiv_1 x'$.  We define $W \defined
        \bigcap_{x' \in F} V_2^{(x',y')}$, which is an open set since it
        is a finite intersection of open sets.  We claim that $V_1
        \times W \subseteq f^{-1}(U)$.  To prove this fact notice that
        if $(a,b) \in V_1 \times W$, then $(a,y) \in V_1 \times V_2$.
        Consider $a' \in F$ such that $a' \equiv_1 a$, we have $(a',y)
        \in V_1 \times V_2$ and therefore $(a',y') \in V_1^{(a',y')}
        \times V_2^{(a',y')} \subseteq f^{-1}(U)$.  In particular, as
        $b \in W \subseteq V_2^{(a',y')}$, we conclude that $(a',b)
        \in f^{-1}(U)$.  Because $a' \equiv_1 a$, this proves that
        $(a,b) \in f^{-1}(U)$.

         Hence, we can build the set $V_1 \times (V_2 \cup W)$
            which strictly contains $V_1 \times V_2$,
            but is still included in $f^{-1}(U)$ which is in contradiction
            with the maximality of $V_1 \times V_2$.

         We have proven that the maximal cylinder sets included in $f^{-1}(U)$
            are $\equiv$-saturated.
            Remark that there can be only finitely many distinct
            $\equiv$-saturated sets in $X_1 \times X_2$.
            Hence, writing $f^{-1}(U)$ as the finite union of maximal
            cylinder sets shows that $f^{-1}(U)$ is a finite
            union of $\equiv$-saturated cylinder sets, which
            allows us to conclude that $f^{-1}(U)$ is compact.

            \medskip
    We have proven that $f$ is a surjective logical map
    and therefore that its image is a \lpps. We can conclude that
    the logical product is a \lpps.

    \medskip
    Let us turn to the statement that $X$ and $X_1\times X_2$ are
    homeomorphic.  The map $f$ is bijective by construction, and we
    already know that it is continuous by definition of a logical
    product, thus it remains to show that $f^{-1}$ is continuous.
    
    Consider a set $K \times X_2$ where $K$ is compact in $X_1$.  The
    set $K$ is then definable using a sentence $\varphi \in
    \FO[\upsigma_1]$ since $X_1$ is a \lpps. Build the sentence $\psi
    \in \FO[\upsigma]$ obtained by relativising the quantifiers in
    $\varphi$ to elements where $\varepsilon_1$ holds.  By
    construction, $\modset{\psi}_{X} = f(K \times X_2)$ and by
    definition of the topology over $X$, $\psi$ is open in $X$.  A
    compact open set of $X_1 \times X_2$ is a finite union of sets of
    the form $K \times X_2$ and $X_1 \times K$ where $K \in
    \CompSat{X_1}$ (resp.\ $K \in \CompSat{X_2})$ thanks
    to~\cref{lem:closure:prodps}.  We conclude that the image
    through~$f$ of a compact open set is a compact open set. In
    particular, $f^{-1}$ is continuous.  Hence $X$ is homeomorphic to
    the product of~$X_1$ and~$X_2$ in the category~$\CatPS$.
\end{proof}

In particular, \cref{lem:closure:prodlps} proves that logical products
provide a concrete representation of products in the category $\CatPS$
as a space of structures. One can wonder why the case of infinite
products, that is seems similar, cannot be considered. The issue in a
generalisation arises from the Feferman-Vaught Theorem, that does not
allow us to handle an equality relation over the infinite product but
only one distinct equality relation per position in the product.
Here is an example where the infinite product is not \lpps.

\begin{example}
    Consider $I = \mathbb{N}$, $X_i \defined \ModF(\upsigma_i)$ with
    $\upsigma_i \defined \emptyset$, and let $\uptau$ be the
    Alexandroff topology for the induced substructure
    ordering~$\subseteq_i$.  Then for any~$i\in I$ and $A,B\in X_i$,
    $A\subseteq B$ if and only if $|A|$ is of cardinality less or
    equal that of~$|B|$.  Hence $(X_i,{\subseteq_i})$ is order
    isomorphic to $(\mathbb N_{>0},{\leq})$ and is a well-quasi-order.
    Therefore, each $(X_i,\Alex{\subseteq_i})$ is Noetherian and each
    $\LPS{X_i}{\uptau}{\FO[\upsigma_i]}$ is thus a \lpps.

    Consider the logical product space~$Z$ of the family $(X_i)_{i \in
      I}$, thus using the signature $\upsigma = \setof{\upepsilon_i}{i
      \in I}$.  We define the $\FO[\upsigma]$ formul\ae\ $\varphi
    \defined \exists x. \exists y. \neg(x = y)$, $\psi_i \defined
    \exists x. \exists y. \varepsilon_i(x) \wedge \varepsilon_i(x)
    \wedge \neg(x = y)$ for $i \in I$ and $\theta_{i,j} \defined
    \exists x. \exists y. \varepsilon_i (x) \wedge \varepsilon_j (x)$
    for $i < j \in I$.

    Notice that
    $\modset{\varphi}_Z = \bigcup_{i \in I} \modset{\psi_i}_Z \cup \bigcup_{i <
        j \in I} \modset{\theta_{i,j}}_Z$, that
    all of the above sentences define open sets in $Z$, but that there
    is no finite subcover.
    Hence there exists a sentence defining a non-compact open set,
    thus $Z$ is not a \lpps.
\end{example}
\end{sendappendix}

    \section{Logical Closure}
    \label{sec:logic}

Consider a set $Z$ equipped with a bounded sublattice
$\mathcal{L}$ of $\parts Z$.  In this section, we provide a way to
consider the closure of a space $X \subseteq Z$ in a suitable topology
so that if $X$ is a \lpps, then its closure also is.  Let us write
$\uptau_{\mathcal{L}} \defined \langle \mathcal{L} \cup
\setof{U^c}{U \in \mathcal{L}} \rangle$ for the topology generated by
the sets of~$\mathcal L$ and their complements.  We call the closure
$\overline{X}$ of $X$ in $(Z, \uptau_{\mathcal{L}})$
its \emph{logical closure}%
.%
\begin{sendappendix}%
Note that the closure of a set is empty if and only if the set itself
is empty, and more generally the following holds.
\begin{fact}
    \label{fact:closure:fmp} For all open sets
    $U \in \uptau_{\mathcal{L}}$, $U \cap X$ is empty if and only if
    $U \cap \overline{X}$ is empty.
\end{fact}
\end{sendappendix}

We show \ifsubmission in \appref{app:sec:logic} \fi that \lpps\ are
stable under logical closures.  For $X\subseteq Z$ and a sublattice
$\mathcal L$ of $\parts Z$, we write $\mathcal L_X\defined \{U\cap
X\mid U\in\mathcal L\}$ for the lattice induced by~$X$.

\begin{restatable}[Stability under logical closure]{proposition}{lemclosurefmplifting}
    \label{lem:closure:fmplifting} Let $X \subseteq
    Y \subseteq \overline{X}$ and $\uptau$ be a topology
    on~$Y$.  If $\LPS{X}{\uptau_X}{\mathcal{L}_X}$ is a \lpps\ for the
    topology $\uptau_X$ induced by~$\uptau$ on~$X$, then so is
    $\LPS{Y}{\uptau}{\mathcal{L}_Y}$.  If $\mathcal{L}'$ is a
    sublattice of $\mathcal{L}$ and $\mathcal{L'}_X$ is a diagram base
    of $X$ then $\mathcal{L'}_Y$ is a diagram base of $Y$.
\end{restatable}

\begin{sendappendix}
    \begin{ifappendix}
        \lemclosurefmplifting*
    \end{ifappendix}
\begin{proof}
    Let $U \in \mathcal{L}$ such that $U \cap Y \in \uptau$. In
    particular, $U \cap Y\in\mathcal L_Y$ is a definable open set of
    $Y$.  By restriction, $U \cap X\in\mathcal L_X$ is a definable
    open set of $X$. Using Alexander's Subbase Lemma, let $(U_i \cap
    Y)_{i \in I}$ be an open cover of~$U$ in~$\uptau$.  By restricting
    to $X$, we can use the fact that $X$ is a {\lpps} to extract a
    finite subset $I_0 \subfin I$ such that $(U_i \cap X)_{i \in I_0}$
    is an open cover of $U \cap X$ in~$\uptau_X$.  In particular, this
    proves that $(U \cap \bigcap_{i \in I_0} U_i^c) \cap X
    = \emptyset$.  Notice that this set is an open set in
    $\uptau_{\mathcal{L}}$, hence we can conclude
    by~\cref{fact:closure:fmp} that
    $U \cap \overline{X} \subseteq \bigcup_{i \in I_0}
    U_i \cap \overline{X}$.  Since
    $Y \subseteq \overline{X}$, $(U_i \cap Y)_{i \in I_0}$
    is an open cover of $U \cap Y$ in~$\uptau$.  Therefore, the definable
    open sets of $Y$ are compact and
    $\LPS{Y}{\uptau}{\mathcal{L}}$ is a \lpps.

    Assume that $\mathcal{L}'_X$ defines a diagram base of $X$.
    Consider a definable open set $U \cap Y\in\mathcal{L}_Y$ of $Y$
    where $U \in \mathcal{L}$.  Remark that $U \cap X\in\mathcal{L}_X$
    is a definable open set in $X$, hence there exists a family
    $(V_i)_{i \in I}$ of elements of $\mathcal{L}'_X$ such that
    $U \cap X = \bigcup_{i \in I} V_i \cap X$.  Since $X$ is a \lpps,
    $U \cap X$ is compact, and therefore $U \cap X = \bigcup_{i \in
    I_0} V_i \cap X$ for some finite $I_0 \subfin I$.  Because
    $\mathcal{L}'_X$ is a lattice, we conclude that $U \cap X = V \cap
    X$ for $V\defined\big(\bigcup_{i \in I_0} V_i\big)$ such
    that $V\cap
    X \in \mathcal{L}'_X$.  Using~\cref{fact:closure:fmp}, this proves
    that $U \cap \overline{X} =
    V \cap \overline{X}$, and because
    $Y \subseteq \overline{X}$ we have $U \cap Y = V \cap
    Y$ where $V\cap Y\in\mathcal{L}'_Y$.  Hence, $\mathcal{L}'_Y$ is a
    diagram base of $Y$.
\end{proof}
\end{sendappendix}

\ifsubmission\subparagraph{Applications of logical closures.}\fi
We now show that~\cref{lem:closure:fmplifting} allows to restate known
preservation theorems and derive new ones.  We consider the case where
$Z = \Mod(\upsigma)$ and $\mathcal{L}
= \modset{\FO[\upsigma]}_{\Mod(\upsigma)}$, and we write $\uptau_\FO$
for the topology~$\uptau_{\mathcal{L}}$.

Let us define $\ModFMP(\upsigma) \subseteq \Mod(\upsigma)$ as the set
of structures whose first-order theory satisfies the \emph{finite
model property}: any definable subset of $\ModFMP(\upsigma)$ has a finite model%
\ifsubmission\ (see~\cref{app:sec:logic})\fi.
We prove that homomorphism preservation can be lifted from
$\ModF(\upsigma)$ (where it holds by \citeauthor{Rossman08}'s Theorem)
to~$\ModFMP(\upsigma)$ in~\cref{cor:closure:fmphpt}.  To our knowledge
this is a new result. \ifsubmission\relax\else\par\fi This follows
from~\cref{lem:closure:fmplifting} and the fact that
$\ModFMP(\upsigma)$ is the closure of $\ModF(\upsigma)$ in the
topology~$\uptau_\FO$\ifsubmission\ (see \appref{app:sec:logic})\fi.
\begin{sendappendix}
\begin{ifappendix}It is easy to see that
$\ModFMP(\upsigma)$ the set
of structures with finite
model property is the closure of $\ModF(\upsigma)$ in the
topology~$\uptau_\FO$.
\end{ifappendix}%
Indeed, consider a structure $A \in \overline{\ModF(\upsigma)}$ and a
sentence $\varphi \in \FO[\upsigma]$ such that $A \models \varphi$.
Because $\modset{\varphi}_{\Mod(\upsigma)}$ is an open set of
$\Mod(\upsigma)$ for~$\uptau_\FO$, this means that
$\modset{\varphi}_{\Mod(\upsigma)} \cap \ModF(\upsigma) \neq
\emptyset$ by definition of the topological closure, hence $\varphi$ has a finite
model.  Conversely, consider a structure $A$ enjoying the finite model
property and let $U$ be a definable open set of $\uptau_\FO$ that
contains $A$. By the finite model property, there exists
$B \in \ModF(\upsigma)$ such that $B \in U$. Hence, $A$ is in the
closure of $\ModF(\upsigma)$.
\end{sendappendix}
\begin{restatable}[Homomorphism preservation for structures with
 the finite model property]{corollary}{corclosurefmphpt}
    \label{cor:closure:fmphpt}
    $\ModFMP(\upsigma)$ has the
    $(\Alex{\homomorphism}, \EPFO[\upsigma])$ preservation property.
\end{restatable}

\begin{sendappendix}
    \begin{ifappendix}
        \corclosurefmphpt*
    \end{ifappendix}
\begin{proof}
    By \citeauthor{Rossman08}'s Theorem~\cite{Rossman08},
    $\ModF(\upsigma)$ has the $(\Alex{\homomorphism}, \EPFO[\upsigma])$
    preservation property.  Since $\ModF(\upsigma)$ is
    downwards-closed for~$\homomorphism$
    inside~$\Mod(\upsigma)$, \cref{cor:prespectral:rhpt} shows that
    the space $\LPS{\ModF(\upsigma)}{\Alex{\to}}{\FO[\upsigma]}$ is a \lpps.
    Moreover, notice that $\ModFMP(\upsigma)
    = \overline{\ModF(\upsigma)}$ and $\EPFO[\upsigma]$ defines a
    diagram base of $\ModF(\upsigma)$.
    Leveraging \cref{lem:closure:fmplifting}, we conclude that
    $\LPS{\ModFMP(\upsigma)}{\Alex{\homomorphism}}{\FO[\upsigma]}$ is
    a {\lpps} and that $\EPFO[\upsigma]$ defines a diagram base of
    $\ModFMP(\upsigma)$.  Now, by \cref{thm:prespectral:rhpt},
    $\ModFMP(\upsigma)$ has the
    $(\Alex{\homomorphism},\EPFO[\upsigma])$ preservation property.
\end{proof}
\end{sendappendix}

Let $\CountableUnions(\upsigma)$ be the set of countable disjoint
unions of finite structures over a finite relational signature
$\upsigma$. We state in~\cref{thm:closure:cufs} another consequence
of \citeauthor{Rossman08}'s Theorem and \cref{lem:closure:fmplifting},
using the fact
$\ModF(\upsigma) \subsetneq \CountableUnions(\upsigma) \subsetneq \ModFMP(\upsigma)
= \overline{\ModF(\upsigma)}$; the same result was first shown
by \citeauthor{Neetil12} in~\cite[Theorem~10.6]{Neetil12}.
\begin{restatable}[Homomorphism preservation for countable unions of finite
    structures]{corollary}{thmclosurecufs}
    \label{thm:closure:cufs}    
    $\CountableUnions(\upsigma)$ has the
    $(\Alex{\homomorphism}, \EPFO[\upsigma])$ preservation property.
\end{restatable}
\begin{sendappendix}
    \begin{ifappendix}
        \thmclosurecufs*
    \end{ifappendix}
\begin{proof}
    Clearly, $\ModF(\upsigma) \subsetneq \CountableUnions(\upsigma)$.
    Regarding
    $\CountableUnions(\upsigma) \subsetneq \ModFMP(\upsigma)$, we
    want to prove
    that for all $A \in \CountableUnions(\upsigma)$ and for all
    $\varphi \in \FO[\upsigma]$, if $A \models \varphi$, then there
    exists a finite structure~$A_\mathrm{fin}$ such that
    $A_\mathrm{fin} \models \varphi$.

    This is an application of the locality of first-order logic.  By
    Gaifman's Locality Theorem~\cite[Theorem
    4.22]{libkin2006locality}, it suffices to prove this assuming
    that~$\varphi$ is a Boolean combination of \emph{basic local
    sentences}, i.e., of sentences of the
    form \begin{equation} \label{eq:local} \exists x_1,\dots
    x_s\mathbin.\big(\bigwedge_{1\leq i\leq
    s}\alpha^{(r)}(x_i)\wedge\bigwedge_{1\leq i< j\leq
    s}d^{>2r}(x_i,x_j)\big), \end{equation} where the
    $\alpha^{(r)}(x_i)$ formula is \emph{$r$-local}
    around~$x_i$~\citep[see e.g.,][Section~4.5]{libkin2012elements}.
    Notice that if a basic local sentence of the form~\eqref{eq:local}
    is satisfied on a structure $B$, then for all structures~$C$, the
    local sentence is also satisfied in the disjoint union $B\uplus
    C$.  Conversely, if a basic local sentence of quantifier rank less
    than~$s$ is satisfied on a structure~$B$, then it is satisfied in
    a union of at most~$s$ connected components of~$B$.  In
    particular, for a structure~$A\in \CountableUnions(\upsigma)$,
    consider the sequence $(A_n)_{n \in \mathbb{N}_{\geq 1}}$ where
    each~$A_n$ is the disjoint union of all components of~$A$ of size
    at most~$n$. For a basic local sentence $\beta$, if
    $A \models \beta$ then $\exists n_0, \forall n \geq n_0,
    A_n \models \beta$.  Conversely, if $A_n \models \beta$ for
    some~$n$, then $A \models \beta$, hence if $A \models \neg \beta$,
    then $\forall n \geq 1, A_n \models \neg \beta$.  We have proven
    that when $\beta$ is a basic local sentence or the negation of a
    basic local sentence, $A \models \beta$ implies $\exists
    n_0, \forall n \geq n_0, A_n \models \beta$.  Notice that this
    property is stable under finite disjunction and finite
    conjunctions: this shows the announced result for Boolean
    combinations of basic local sentences.

    As a consequence,
    $\ModF(\upsigma) \subsetneq \CountableUnions(\upsigma) \subsetneq \overline{\ModF(\upsigma)}$
    and we apply the same reasoning as in~\cref{cor:closure:fmphpt}.
\end{proof}
\end{sendappendix}

    \section{Projective Systems}
    \label{sec:proj}
    A natural construction in the category of topological spaces is
the \emph{projective limit}, and the category $\CatSpec$ of spectral
spaces and spectral maps is closed under this
construction~\cite[Corollary 2.3.8]{spectral2019}.  As an
illustration, we show in \cref{sub:hpt} that
$\LPS{\ModF(\upsigma)}{\Alex{\homomorphism}}{\FO[\upsigma]}$ is the
projective limit of a system of Noetherian spaces, which provides an
alternative understanding of Rossman's Theorem~\cite{Rossman08}.  In
fact, as we show in \cref{sub:completeness}, any pre-spectral space is
the limit of a projective system of Noetherian spaces.

\subsection{Projective Systems}
A \emph{projective system} $\mathcal{F}$ in a category $\mathbf{C}$
assigns to each element $i$ of a directed partially ordered set~$I$
an object $X_i$ and to each ordered pair $i \leq j$ a
so-called \emph{bonding map} $f_{i,j} \colon X_i \to
X_j$
so that, for all $i,j,k \in I$ with $k \leq j \leq i$, we have
$f_{i,i} = \mathrm{id}_{X_i}$ and $f_{j,k} \circ f_{i,j} = f_{i,k}$.
The \emph{projective limit} of a projective system $\mathcal{F}$
is an object~$X$ with maps~$f_i \colon X \to X_i$
\emph{compatible} with the system $\mathcal{F}$, which means that,
for all $i \geq j$, $f_{i,j} \circ f_i = f_j$.  Moreover, $X$
satisfies a universal property: whenever $\{ g_i \colon Y \to
X_i\}_{i \in I}$ is a family of maps compatible with~$\mathcal{F}$, there exists a unique map $g \colon Y \to X$ such that
$g_i = f_i \circ g$ for all $i \in I$.

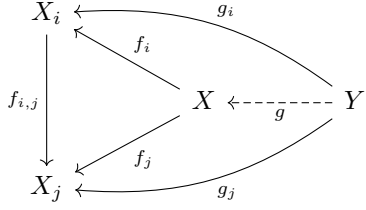
\begin{figure}[bt]
    \centering
    \begin{tikzcd}[column sep=4em]
        X_{i} \arrow[dd,swap,"f_{i,j}"] & & \\
                                        & X \arrow[lu,swap,"f_i"] 
                                            \arrow[ld,"f_j"]
                                        & Y \arrow[l,"g",dashed]
                                            \arrow[llu,swap,"g_i",bend
                                            right=20]
                                            \arrow[lld,"g_j",bend
                                            left=20]
                                        & \\
        X_{j} & & \\
    \end{tikzcd}
    \caption{The commutative diagram of a projective system.}
    \label{fig:hpt:projsystem}
\end{figure}
\begin{ifappendix}
    Unfortunately, there exists projective systems in $\CatPS$
    that do not have limits, as is witnessed by~\cref{ex:hpt:scott}
    which is a slight adaptation of~\cite[Example~3]{STONE1979203}.
\end{ifappendix}%
\begin{sendappendix}%
    \begin{ifappendix}\subparagraph{Existence of Limits.}\end{ifappendix}
        Recall that we write $\CatPS$ for the category of pre-spectral spaces
        and spectral maps.  The category~$\CatSpec$ of spectral spaces and
        spectral maps is closed under projective
        systems~\cite[Corollary 2.3.8]{spectral2019}, meaning
        that every projective system in $\CatSpec$ has a limit in~$\CatSpec$.
        This is not the case for $\CatPS$, as can be seen by the
        following adaptation of~\cite[Example~3]{STONE1979203}.%

\begin{example}\label{ex:hpt:scott}
    Consider as in \cref{ex:related:cycles} the spaces
    $(\Cycles,\uptau_n)$ of finite cycles with the topology generated
    using sentences having models of size at most $n$ or finitely many
    counter-models (when $n = 0$, this is the co-finite topology).
    Recall that $(\Cycles,\uptau_n)$ is a Noetherian space for all
    $0 \leq n < \infty$, and is therefore pre-spectral.  Observe that
    the identity maps between $(\Cycles,\uptau_i)$ and
    $(\Cycles,\uptau_j)$ are defining a projective system in~$\CatPS$.
    Assume that $\{f_i : X \to (\Cycles, \uptau_i) \}_{i \in I}$
    is the limit of this projective system in $\CatPS$.
    The commutation property $id_{i,j} \circ f_i = f_j$
    shows that all $f_i = f_j$ for $i,j \in I^2$
    and we let $f \defined f_i$ for some $i \in I$. 
    The function $f$ is continuous
    when endowing $\Cycles$ with the discrete topology.
    Indeed, consider $C \in \Cycles$ a given cycle,
    then $\{C\}$ is open in $(\Cycles, \uptau_{|C|})$ and therefore
    $f^{-1}(\{C\})$ is open in $X$. Since $X$ is a pre-spectral space,
    it is compact, hence its image through a continuous function
    is compact, but $\Cycles$ with the discrete topology is not
    compact which is a contradiction.
\end{example}
\end{sendappendix}%
Let us introduce here the category of topological spaces and
continuous maps, denoted by~$\CatTop$.  A projective system
in~$\CatPS$ is a projective system of topological spaces
in~$\CatTop$. A projective system in~$\CatPS$ always has a limit when
considered as a projective system in~$\CatTop$; we give a sufficient
condition for this space to be the limit in $\CatPS$\ifsubmission\
(see \cref{app:sec:proj})\fi.

\begin{restatable}[Transfer of projective limits]{lemma}{lemhptlimit}
    \label{lem:hpt:limit}
    Let $\mathcal{F}$ be a projective system of pre-spectral spaces
    in~$\CatPS$.
    If $\{f_i \colon X \to X_i\}_{i \in I}$ 
    is the limit of~$\mathcal{F}$
    in~$\CatTop$ where
    the maps $f_{i} \colon X \to X_i$
    are spectral,
    then it is the limit of $\mathcal{F}$ in~$\CatPS$.
    Moreover, $\CompSat{X} = \bigcup_{i \in I} \setof{f_i^{-1}(V)}{V \in
    \CompSat{X_i}}$.
\end{restatable}

\begin{sendappendix}%
    \begin{ifappendix}%
    \subparagraph{Proof of \cref{lem:hpt:limit}.} Recall that
    $\CatTop$ denotes the category of topological spaces and
    continuous maps.
        \lemhptlimit*
    \end{ifappendix}
\begin{proof}
    Since $X$ is the projective limit in $\CatTop$, we can assume that
    it is the standard projective limit, that is
    $X \defined \setof{\vec{x} \in \prod_{i \in I} X_i}{\forall i \leq
    j \in I, x_j = f_{i,j} (x_i)}$ with the topology inherited from
    $\prod_{i \in I} X_i$, and that $f_i$ is the projection over
    component~$i$, that is $f_i( \vec{x}) = x_i$~\cite[Exemple
    4.12.8]{goubault2013non}.
    We will now prove that the set of compact-open
    sets of~$X$ is a bounded sublattice generating the topology and
    thus that~$X$ is pre-spectral.

    Consider an open set $U$ of $X$: by definition of the topology of
    $X$, it is a union $\bigcup_{i \in I} f_i^{-1}(U_i)$ where
    each~$U_i$ for $i \in I$ is open in~$X_i$.  For every $i \in I$,
    $X_i$ is pre-spectral, meaning that $\CompSat{X_i}$ generates the
    topology of~$X_i$, thus each $U_i$ can be written as a union of
    compact opens of~$X_i$. Note that for all~$i \in I$ and all
    compact open subsets~$K$ of $X_i$, $f_i^{-1} (K)$ is a compact
    open set in~$X$ since~$f_i$ is spectral.  Therefore, $U$ itself
    can be written as a union of pre-images~$f_i^{-1}(K)$,
    which are compact open
    in~$X$.

    Consider now a compact open set~$U$ of~$X$. We have just seen that
    it is a union of compact sets of~$X$, and since~$U$ is compact, we
    can extract a finite cover, so $U$ is a finite union of compact
    open sets of the form $f_{i_j}^{-1}(K_j)$ for $1 \leq j \leq n$
    where $\forall 1 \leq j \leq n, i_j \in I$ and $\forall 1 \leq
    j \leq n, K_j \in \CompSat{X_{i_j}}$.  Since~$I$ is directed,
    there is an index~$i$ above all~$i_j$ for $1 \leq j \leq n$.  Let
    $K_j' \defined f_{i, i_j}^{-1}(K_j)$ for $1 \leq j \leq n$; it is
    a compact open of~$X_i$ because~$f_{i, i_j}$ is
    spectral.  Notice that $f_{i,i_j} \circ f_i = f_{i_j}$ hence
    $f_{i}^{-1}(K_j') = f_{i_j}^{-1}(K_j)$.  Hence $U = f_i^{-1}
    (\bigcup_{j = 1}^n K_j')$ where $\bigcup_{j = 1}^n
    K_j'\in\CompSat{X_i}$ since~$X_i$ is prespectral and
    therefore~$\CompSat{X_i}$ is a lattice.  This shows that
    $\CompSat{X} = \setof{f_i^{-1}(K)}{\forall i \in I, \forall
    K \in \CompSat{X_i}}$, and that $\CompSat{X}$ generates the
    topology of~$X$.

    Consider two compact open sets of $X$, $U$ and $U'$.  They are
    obtained as pre-images $U = f_i^{-1} (V)$ and $U' = f_j^{-1}(V')$
    with $V\subseteq X_i$ and $V'\subseteq X_j$ compact open.  Because
    $I$ is directed, there exists $k\in I$ such that $k \geq i,j$.
    Because $f_{j} \circ f_{k,j} = f_k$, we can see that $U = f_k^{-1}
    (f_{k,i}^{-1}(V))$ and $U' = f_k^{-1}(f_{k,j}^{-1}(V'))$. Because
    the maps are morphisms and compact open sets of the pre-spectral
    space $X_k$ form a lattice, we can see that their intersection and
    union form compact open sets of~$X_k$, and therefore compact open
    sets are stable under binary intersection and binary unions
    in~$X$.  This shows that $\CompSat{X}$ is indeed a lattice.  It
    remains to prove that it is a bounded one.
    Since~$I$ is non empty, let us pick~$i \in I$ and notice that
    $f_i^{-1}(X_i) = X$ since~$X_i$ is compact and~$f_i$ is spectral:
    this proves that~$X$ is compact.  Hence $\CompSat{X}$ is a bounded
    sublattice of $\parts X$.

    We finally prove that~$X$ is the limit in~$\CatPS$.  Consider $\{
    g_i \colon Y \to X_i \}_{i \in I}$ any family of spectral maps
    compatible with~$\mathcal{F}$, namely such that $g_j \circ f_{i,j}
    = g_i$ for all $i \leq j \in I$.  In particular, the maps~$g_i$
    are continuous.  By the universal property of~$X$ as a limit in
    $\CatTop$, there exists a unique map $g \colon Y \to X$ such that
    $f_i \circ g = g_i$ for every~$i \in I$.  In particular, consider
    a compact open set of~$X$: it can written as~$f_i^{-1}(K)$ for
    some~$i \in I$ and~$K \in \CompSat{X}$, hence $g^{-1}(f_i^{-1}(K))
    = g_i^{-1}(K)$ is compact because~$g_i$ is spectral.  Hence, $g$
    is spectral as well.  The uniqueness of~$g$ follows from the
    uniqueness in~$\CatTop$.  This shows that $\{ f_i \colon X \to
    X_i \}_{i \in I}$ is the limit of~$\mathcal{F}$ in~$\CatPS$.
\end{proof}

\end{sendappendix}

\subsection{Application to the Homomorphism Preservation Theorem}
\label{sub:hpt}

Throughout this section, we fix a finite relational signature
$\upsigma$ and a downwards-closed subset $X$ of $\ModF(\upsigma)$ for
the homomorphism ordering~$\homomorphism$, i.e., $X$ is
co-homomorphism closed. We will see
how \citeauthor{Rossman08}'s Theorem can be explained as the existence
of a projective limit.

\subparagraph{\ifsubmission{\boldmath $n$}\else$n$\fi-Homomorphisms.}
Let us define the \emph{tree-depth} $\td(A)$ of a finite structure~$A$
as the tree-depth~$\td(\mathcal G(A))$ of its associated \emph{Gaifman
graph}~$\mathcal G(A)$~\cite[Definition~4.1]{libkin2012elements}.
Following the idea of the original proof by
\citet[Section~3.2]{Rossman08}, we are going to use
quasi-orders that are \emph{coarser} than the homomorphism
quasi-order, and refine those progressively.  For every
$n \in \mathbb{N}$, we define $A \homomorphism_n B$ if for every
structure $C$ of tree-depth at most~$n$, $C
\homomorphism A$ implies $C \homomorphism B$.  Note that on finite
structures, $A \homomorphism B \iff A \homomorphism_{\td(A)} B$.
Then the intersection of all
 the $\homomorphism_n$ relations is~$\homomorphism$.
Let us consider
the corresponding Alexandroff topologies:
$X \defined
\LPS{X}{\Alex{\to}}{\FO[\upsigma]}$ and for $n
\in \mathbb{N}$, let $X_n \defined
\LPS{X}{\Alex{\to_n}}{\FO[\upsigma]}$.

\begin{sendappendix}
\begin{ifappendix}\subparagraph{\ifsubmission{\boldmath
$n$}\else$n$\fi-Cores and Homomorphism Preservation.}~\!\!\unskip
\end{ifappendix}
For $A \in \ModF(\upsigma)$ and $n \geq 1$, there exists a structure
$\NCore{n}{A}$ of tree-depth at most~$n$ such that
$A \homomorphism_n \NCore{n}{A}$, $\NCore{n}{A} \homomorphism_n A$,
and furthermore $A \homomorphism_n B$ if and only if $\NCore{n}{A} \to
B$~\cite[definitions~3.6 and~3.10 and Lemma~3.11]{Rossman08}.  Notice that for
all $n \geq 1$, if $A \in X$ then $\NCore{n}{A} \in X$ since $X$ is
downwards closed.
\end{sendappendix}

\subparagraph{\citeauthor{Rossman08}'s Lemma.}
In his paper~\cite{Rossman08}, \citeauthor{Rossman08} provides a
function $\rho\colon\mathbb N\to\mathbb N$ and relates
indistinguishability in the fragment $\FO_n[\upsigma]$ of first-order
logic with at most~$n$ quantifier alternations with
${\rho(n)}$-homomorphism equivalence~\cite[Corollary~5.14]{Rossman08}.
We state this result in a self-contained manner below~\citep[see
also][Theorem~10.5]{Neetil12}.

\begin{restatable}[\protect{\citeauthor{Rossman08}'s Lemma~\cite{Rossman08}}]{lemma}{thmhptrosslem}
    \label{thm:hpt:rosslem} There exists $\rho\colon\mathbb
    N\to\mathbb N$ such that, for all $n\in\mathbb N$, if
    $\varphi \in \FO_n[\upsigma]$ is closed under homomorphisms, then it is
    closed under $\rho(n)$-homomorphisms.
\end{restatable}
\begin{sendappendix}
    \begin{ifappendix}\thmhptrosslem*\end{ifappendix}
    \begin{proof}%
    We prove that the actual statement of \citeauthor{Rossman08}'s
    Lemma in~\cite[Corollary 5.14]{Rossman08}
    implies~\cref{thm:hpt:rosslem}.
    Corollary~5.14 in~\citep{Rossman08} states that for every pair
    $A,B \in \ModF(\upsigma)$ such that $A \homomorphism_{\rho(n)} B$
    and $B \homomorphism_{\rho(n)} A$ there exists two finite
    structures $\widetilde A$ homomorphically equivalent to $A$ and
    $\widetilde B$ homomorphically equivalent to $B$ (more precisely,
    $A$ and~$B$ are \emph{retracts} of~$\widetilde A$ and~$\widetilde
    B$), such that $\widetilde A$ and $\widetilde B$ satisfy the
    same~$\FO_n[\upsigma]$ sentences.

    Assume now that $\varphi \in \FO_n[\upsigma]$ is closed under
    homomorphisms, $A\models\varphi$, and $A \homomorphism_{\rho(n)}
    C$. Let us define $B\defined \NCore{\rho(n)}{A}$:
    then~$A\homomorphism_{\rho(n)}B$, $B\homomorphism_{\rho(n)}A$ and
    furthermore $B \homomorphism C$.  Apply~\cite[Corollary
    5.14]{Rossman08} to~$A$ and~$B$. Then $\widetilde
    A\models \varphi$ because $A\homomorphism\widetilde A$ and
    $\varphi$ is closed under homomorphisms. Therefore $\widetilde
    B \models \varphi$ because $\widetilde A$ and $\widetilde B$ are
    $n$-elementary equivalent. Finally, $\widetilde B\homomorphism
    B\homomorphism C$, thus $C \models \varphi$ because $\varphi$ is
    closed under homomorphisms.  This shows that~$\varphi$ is closed
    under $\rho(n)$-homomorphisms.
\end{proof}
\end{sendappendix}
\Citeauthor{Rossman08}'s Lemma is the combinatorial heart
of \citeauthor{Rossman08}'s Theorem, so the developments in this
section are really meant to show how the pre-spectral framework can
capture \citeauthor{Rossman08}'s arguments translating the technical
statement from \cref{thm:hpt:rosslem} into a proof of homomorphism
preservation in the finite.

\subparagraph{Projective System.}  We are now ready obtain $X$
as a limit of a projective system in~$\CatPS$.
We are going to exploit \cref{thm:hpt:rosslem} through the definition
of the topological spaces
$Y_n \defined \LPS{X}{\Alex{\homomorphism}}{\FO_n[\upsigma]}$ for all~$n$.
We will use the following consequence of \citeauthor{Rossman08}'s Lemma\ifsubmission\ (see \cref{app:sec:proj})\fi.
\begin{restatable}{claim}{facthptprescmpct}
    \label{fact:hpt:prescmpct}
    $\forall n \geq 1, \CompSat{Y_n} \subseteq \CompSat{X_{\rho(n)}} \subseteq \CompSat{X}$.
\end{restatable}

\begin{sendappendix}

   \begin{ifappendix}
        \facthptprescmpct*
    \end{ifappendix}
\begin{claimproof}%
    The proof is split in three parts: we first show that $X_m$ is
    Noetherian, then that a compact open
    set of~$X_m$ is compact open in~$X$, and finally that a compact
    open in~$Y_n$ is a compact open in~$X_{\rho(n)}$.

    For the first step, let~$U$ be an open subset of~$X_m$.
    By \cite[Lemma~3.11]{Rossman08}, for $A,B \in X$,
    $A \homomorphism_m B \iff \NCore{m}{A} \homomorphism B$.  Consider
    the associated set of cores $F\defined\{\NCore{m}{A}\mid A\in
    U\}$: then $U$ is the upward closure of~$F$ under homomorphisms.
    Furthermore, $F$ is finite up to homomorphic equivalence
    by~\cite[Lemma~3.9]{Rossman08}, and thus also finite up to
    $m$-homomorphic equivalence.  Finally, still
    by \cite[Lemma~3.11]{Rossman08}, $U$ is also the upward closure
    of~$F$ under $m$-homomorphisms.  By~\cref{rk:alex:compact}, $U$ is
    thus compact in~$X_m$.  This shows that~$X_m$ is Noetherian.

    As we have seen, any open subset~$U$ of~$X_m$ is the upward
    closure for~$\homomorphism$ of a finite set~$F$ of cores.
    Thus~$U$ is a compact open set in~$X$ by \cref{rk:alex:compact}.
    We have proven that a compact open set of~$X_m$ is also compact
    open in~$X$.

    Finally, consider a compact open set $U$ of
    ${Y_n}$. By~\cref{thm:hpt:rosslem}, $U$ is open in~$X_{\rho(n)}$
    as well. Furthermore, $U$ is compact in~$X_{\rho(n)}$ since
    $X_{\rho(n)}$ is Noetherian.
\end{claimproof}

\end{sendappendix}

The following theorem was famously first shown by \citeauthor{Rossman08}
in~\cite[Corollary~7.1]{Rossman08}.  A more recent proof by
\citet{rossman16} uses lower bounds from circuit complexity.  Similar
results were shown by~\citet[Section~10.7]{Neetil12} when assuming
essentially the same statement as \cref{thm:hpt:rosslem}; in fact,
carefully unwrapping the hypotheses of the \emph{topological
preservation theorem} of \citet[Theorem~10.3]{Neetil12} leads to the
very definition of a projective system.
\begin{theorem}
    \label{thm:hpt:hpt}
    Let $\upsigma$ be a finite relational
    signature and $X$ be a non-empty downwards-closed subset of $\ModF(\upsigma)$
    for~$\homomorphism$. Then~$X$ has the $(\Alex{\homomorphism},
    \EPFO[\upsigma])$ preservation property.
\end{theorem}
\begin{proof}
    Consider the projective system
    $\mathcal{F} \defined \{ \mathrm{id}_{i,j} \colon Y_i \to
    Y_j\}_{i\leq j\in I}$ indexed by
    $I\defined\mathbb{N}\setminus\{0\}$. Each space $Y_i$ is
    Noetherian for all $i\in I$ because $\FO_i[\upsigma]$ contains
    finitely many non-equivalent sentences, hence $Y_i$ contains
    finitely many open sets. Hence
    $\CompSat{Y_i}=\Alex{\homomorphism}\cap\modset{\FO_i[\upsigma]}_X$. Also,
    the maps $\mathrm{id}_{i,j}$ are spectral and~$\mathcal F$ is a
    projective system in $\CatPS$.  \Cref{fact:hpt:prescmpct} shows
    that the identity map $\mathrm{id}_{i} : X \to Y_i$ is a spectral
    map for all $i\in I$.

    Assume that $\{g_i \colon Z \to Y_i \}_{i \in I}$ is a collection
    of morphisms in $\CatTop$ such that $\forall i \geq j \in I, g_j
    = \mathrm{id}_{i,j} \circ g_i$.  Since $\mathrm{id}_{i,j}$ is the
    identity map, all the maps $(g_i)_{i \in I}$ are equal.  In
    particular, one can build $g \colon Z \to X$ defined by any one of
    them.  Let us show that~$g$ is a continuous map.  If~$U$ is a
    definable open set of~$X$, then~$U$ is a definable open set in
    $Y_n$ for some~$n$, hence $g^{-1}(U) = g_n^{-1}(U)$ is open.
    Since~$X$ has a base of definable open sets, this proves that~$g$
    is continuous.

    Assume that~$g'$ is an other continuous map making the diagram
    commute. As $I$ is non empty, consider some~$i \in I$, we have
    $g_i = \mathrm{id}_i \circ g = \mathrm{id}_i \circ g'$.
    Since~$f_i$ is the identity map we conclude~$g = g'$.

    We have shown that $X$ is the limit of $\mathcal{F}$
    in~$\CatTop$. Since the maps $\mathrm{id}_i \colon X \to X_i$
    are spectral, \cref{lem:hpt:limit}
    shows that~$X$ is a pre-spectral space
    such that $\CompSat{X} = \bigcup_{i \in I} \CompSat{Y_i}
    = \Alex{\homomorphism} \cap \modset{\FO[\upsigma]}_X$.
    In particular, $X$ is a \lpps.
    As $X$ is downwards-closed,
    by~\cref{cor:prespectral:rhpt} it has the
    $(\Alex{\homomorphism}, \EPFO[\upsigma])$-preservation property.
\end{proof}

\subsection{Completeness}
\label{sub:completeness}

We are now going to prove that any pre-spectral space can be obtained
as a solution to a projective system of pre-spectral spaces, showing
that the proof method of the previous sub-section is in some sense
complete. In fact, this system is going to contain only Noetherian
spaces\ifsubmission\ (see \appref{app:sec:proj})\fi.
It is analoguous to the fact that any
spectral space is a projective
limit of finite~$T_0$ spaces~\cite[Proposition~10]{hochster1969prime}.

\begin{restatable}[Pre-spectral spaces are limits of Noetherian spaces]{proposition}{lemhptcompleteness}
    \label{lem:hpt:completeness}
    Let $(X,\uptau)$ be a pre-spectral space,
    there exists a projective system of Noetherian spaces
    in $\CatPS$ such that $X$ is the limit of
    this projective system.
\end{restatable}

\begin{sendappendix}

  \begin{ifappendix}
    \subparagraph{Pre-spectral Spaces are Limits.}
    We provide here the proof of \cref{lem:hpt:completeness}.
    \lemhptcompleteness*
    \end{ifappendix}
\begin{proof}
    We index our projective system by the finite subsets
    of~$\CompSat{X}$ ordered by inclusion. Whenever $K \subfin
    \CompSat{X}$, let us define $X_K \defined (X, \langle K \rangle)$,
    that is, $X$ with the topology generated by the finite collection
    $K$ of compact open sets of $X$.  Note that $X_K$ is Noetherian
    and that the maps $\mathrm{id}_{K,K'} \colon X_K \to
    X_{K'}$ are spectral maps whenever $K' \subseteq K$, so this is a
    projective system in~$\CatPS$.  Moreover,
    $\mathrm{id}_K \colon X \to X_K$ is also spectral.

    It remains to check that~$X$ is the limit.  Assume that $Y$ is a
    solution to the projective system in $\CatPS$: there exists
    spectral maps $g_K \colon Y \to X_K$ for all $K \subfin
    \CompSat{X}$.  Remark that $\mathrm{id}_{K,K'} \circ g_K = g_{K'}$
    by definition, hence $g_K = g_{K'}$.
    Define $g \colon Y \to X$ as $g(y) \defined
    g_{\emptyset} (y)$; notice that $g$ is spectral as well.
    Finally, $g$
    is unique, as an other map $h \colon Y \to X$
    making the diagram commute satisfies
    $g_{\emptyset} = f_{\emptyset} \circ g = f_{\emptyset} \circ h$,
    hence $g = h$.
\end{proof}

\Cref{lem:hpt:completeness} should be contrasted with
\cref{ex:hpt:scott}, which showed that the projective limit of
Noetherian spaces can be a non pre-spectral space.  The existence of
preservation theorems can therefore be reduced to the existence of a
certain projective limit.

\end{sendappendix}

    \section{Concluding Remarks}
    In this paper, we have introduced a general framework for preservation
results, mixing topological and model-theoretic notions.  The key
notion here is the one of \emph{logically presented pre-spectral
spaces}, which requires the (topological) compactness of the definable
sets of interest.  This definition captures simultaneously the
classical proofs of preservation theorems over the class of all
structures (we detailed the case of the \ltrsk\ in \cref{sub:ltrsk})
and all the known preservation results over classes of finite
structures in the literature (see \cref{cor:prespectral:rhpt}).  Our
approach is comparable to the one adopted in the \emph{topological
preservation theorem} of \citet[Theorem~10.3]{Neetil12}, in that we
employ topological concepts to present a generic preservation theorem;
however we believe our formulation to be considerably simpler and more
flexible. %

We have developed a mathematical toolbox for working with logically
presented pre-spectral spaces, allowing to build new spaces from known
ones.  Besides relatively mundane stability properties under suitable
notions of morphisms, subspaces, finite sums, and finite
products---which still required quite some care in order to account
for first-order definability\mbox{---,} we have shown that more exotic
constructions through topological closures or projective limits of
topological spaces could also be employed.  Those last two
constructions give an alternative viewpoint
on \citeauthor{Rossman08}'s proof of homomorphism preservation over
the class of finite structures (\cref{thm:hpt:hpt}), and a new
homomorphism preservation result over the class of structures with
the finite model property (\cref{cor:closure:fmphpt}).

    \section*{Acknowledgements}
    I thank Jean Goubault-Larrecq and Sylvain Schmitz
for their help and support in writing this paper.

    \bibliographystyle{plainnat}

\end{document}